\documentclass[useAMS,usenatbib,letterpaper]{mn2e}
\usepackage{graphicx}
\usepackage{supertabular,wrapfig,placeins}
\usepackage{graphicx,amsmath,amssymb,subfigure,multirow}
\usepackage{marvosym, times}
\usepackage{graphics}
\usepackage{natbib}
\usepackage{amssymb}
\usepackage{amsmath}
\usepackage{euscript}
\usepackage{setspace}
\usepackage{fancyhdr}
\usepackage{lscape}
\usepackage{afterpage,rotating,url}
\usepackage{xcolor}

\def\simgt{\mathrel{\lower0.6ex\hbox{$\buildrel {\textstyle >}
 \over {\scriptstyle \sim}$}}}
\def\simlt{\mathrel{\lower0.6ex\hbox{$\buildrel {\textstyle <}
 \over {\scriptstyle \sim}$}}}

\newcommand{\mnras}{MNRAS}
\newcommand{\apjs}{ApJS}
\newcommand{\apj}{ApJ}
\newcommand{\aap}{A\&A}
\newcommand{\aaps}{A\&AS}
\newcommand{\pasa}{PASA}
\newcommand{\apjl}{ApJL}
\newcommand{\araa}{ARA\&A}
\newcommand{\aj}{AJ}

\newcommand{\pasp}{PASP}

\newcommand{\trm}{\textrm}

\begin{document}

\date{Accepted 2016 January 20. Received 2016 January 19; in original form 2015 August 20.}

\title[A pilot multi-epoch continuum survey with ASKAP-BETA]{Wide-field broadband radio imaging with phased array feeds:\\a pilot multi-epoch continuum survey with ASKAP-BETA}

\author[Heywood et al.]
{\parbox{\textwidth}{
\begin{flushleft}
I.~Heywood$^{1,2}$\thanks{Email: {\tt ian.heywood@csiro.au}}, 
K.~W.~Bannister$^{1,3}$, 
J.~Marvil$^{1}$, 
J.~R.~Allison$^{1,3}$, 
L.~Ball$^{1}$, 
M.~E.~Bell$^{1}$, 
D.~C.-J.~Bock$^{1}$,
M.~Brothers$^{1}$,
J.~D.~Bunton$^{1}$,
A.~P.~Chippendale$^{1}$,
F.~Cooray$^{1,4}$,
T.~J.~Cornwell$^{1,5}$
D.~DeBoer$^{1,6}$,
P.~Edwards$^{1}$,
R.~Gough$^{1}$, 
N.~Gupta$^{1,7}$,
L. Harvey-Smith$^{1}$,
S.~Hay$^{1}$,
A.~W.~Hotan$^{1}$,
B.~Indermuehle$^{1}$,
C.~Jacka$^{1}$,
C.~A.~Jackson$^{1,8,9}$,
S.~Johnston$^{1}$,
A.~E.~Kimball$^{1}$,
B.~S.~Koribalski$^{1}$,
E.~Lenc$^{1,9,10}$,
A.~Macleod$^{1}$,
N.~McClure-Griffiths$^{1,11}$,
D.~McConnell$^{1}$,
P.~Mirtschin$^{1}$,
T.~Murphy$^{9,10}$,
S.~Neuhold$^{1}$,
R.~P.~Norris$^{1}$,
S.~Pearce$^{1}$,
A.~Popping$^{1,8,9}$,
R.~Y.~Qiao$^{1,12}$,
J.~E.~Reynolds$^{1}$,
E.~M.~Sadler$^{9,10}$,
R.~J.~Sault$^{1,13}$,
A.~E.~T.~Schinckel$^{1}$,
P.~Serra$^{1}$,
T.~W.~Shimwell$^{1,14}$,
J.~Stevens$^{1}$,
J.~Tuthill$^{1}$,
A.~Tzioumis$^{1}$,
M.~A.~Voronkov$^{1}$,
T.~Westmeier$^{1,9}$, 
M.~T.~Whiting$^{1}$\\
\end{flushleft}
}
\footnotesize
\\
$^{1}$CSIRO Astronomy and Space Science, Australia Telescope National Facility, P.O. Box 76, Epping, NSW 1710, Australia\\
$^{2}$Department of Physics and Electronics, Rhodes University, P.O. Box 94, Grahamstown, 6140, South Africa\\
$^{3}$Bolton Fellow\\
$^{4}$1 -- 7 Rowe Street, Eastwood, NSW 2122, Australia\\
$^{5}$Tim Cornwell Consulting\\
$^{6}$Radio Astronomy Laboratory, University of California Berkeley, 501 Campbell, Berkeley, CA 94720--3411, USA\\
$^{7}$Inter-University Centre for Astronomy and Astrophysics, Post Bag 4, Ganeshkhind, Pune University Campus, Pune 411 007, India\\
$^{8}$International Centre for Radio Astronomy Research (ICRAR), University of Western Australia, 35 Stirling Highway, Crawley, WA 6009, Australia\\
$^{9}$ARC Centre of Excellence for All-sky Astrophysics (CAASTRO)\\
$^{10}$Sydney Institute for Astronomy, School of Physics, University of Sydney, NSW 2006, Australia\\
$^{11}$Research School of Astronomy \& Astrophysics, The Australian National University, Canberra ACT 2611, Australia\\
$^{12}$Sonartech ATLAS Pty Ltd, Unit G01, 16 Giffnock Avenue, Macquarie Park  NSW  2113\\
$^{13}$School of Physics, University of Melbourne, VIC 3010, Australia\\
$^{14}$Leiden Observatory, Leiden University, PO Box 9513, NL-2300 RA, Leiden, The Netherlands}

\maketitle

\clearpage

\begin{abstract}

The Boolardy Engineering Test Array is a 6~$\times$~12 m dish interferometer and the prototype of the Australian Square Kilometre Array Pathfinder (ASKAP), equipped with the first generation of ASKAP's phased array feed (PAF) receivers. These facilitate rapid wide-area imaging via the deployment of simultaneous multiple beams within a $\sim$30 square degree field of view. By cycling the array through 12 interleaved pointing positions and using 9 digitally formed beams we effectively mimic a traditional 1 hour~$\times$~108 pointing survey, covering $\sim$150 square degrees over 711 -- 1015 MHz in 12 hours of observing time. Three such observations were executed over the course of a week. We verify the full bandwidth continuum imaging performance and stability of the system via self-consistency checks and comparisons to existing radio data. The combined three epoch image has arcminute resolution and a 1$\sigma$ thermal noise level of 375 $\mu$Jy beam$^{-1}$, although the effective noise is a factor $\sim$3 higher due to residual sidelobe confusion. From this we derive a catalogue of 3,722 discrete radio components, using the 35\% fractional bandwidth to measure in-band spectral indices for 1,037 of them. A search for transient events reveals one significantly variable source within the survey area. The survey covers approximately two-thirds of the \emph{Spitzer} South Pole Telescope Deep Field. This pilot project demonstrates the viability and potential of using PAFs to rapidly and accurately survey the sky at radio wavelengths.

\end{abstract}


\section{Introduction}
\label{sec:intro}

Continuum observations at radio wavelengths have been a key component of observational astrophysics for many decades.
The fact that radio observations are not affected by dust obscuration means that they lack many selection biases that exist in observations at other wavebands, and a typical source at the bright end of the radio luminosity function will be associated with a radio-loud active galactic nucleus (AGN) with a median cosmological redshift of $\sim$1 (Condon, 1984). Moving towards fainter flux limits, radio observations become sensitive to the radio quiet AGN population, and an increasing fraction of galaxies whose radio synchrotron emission is driven by star formation (Condon, 1992). Radio continuum observations thus provide unique insight into black hole activity (e.g.~Jarvis \& Rawlings, 2000; Smol{\v c}i{\'c} et al., 2009b; Rigby et al., 2011; McAlpine, Jarvis \& Bonfield, 2013; Banfield et al., 2014; Best et al., 2014) and star formation (e.g.~Seymour et al., 2008, Smol{\v c}i{\'c} et al., 2009a; Jarvis et al., 2015a) across the history of the Universe.

Sky surveys typically have a trade-off between depth and area. Radio surveys with the broadest coverage at $\sim$gigahertz frequencies tend to be `flagship' projects, occupying a significant fraction of available telescope time and covering most of the entire visible sky by means of a very large number of short snapshot pointings to $\sim$mJy beam$^{-1}$ depths. Examples include the NRAO VLA (Very Large Array) Sky Survey (NVSS; Condon et al., 1998), Faint Images of the Radio Sky at Twenty-cm (FIRST; Becker, White \& Helfand, 1995), and the Sydney University Molonglo Sky Survey (SUMSS; Bock, Large \& Sadler, 1998; Mauch et al., 2003). The very deepest observations tend to cover only a single primary beam of the instrument, for example the Lockman Hole observation of Condon et al.~(2012) which reaches a depth of approximately 1 $\mu$Jy beam$^{-1}$. There are many examples that sit somewhere between these two extremes that typically cover a few square degrees over numerous extragalactic deep fields, where radio observations form one part of a panchromatic picture. Such surveys have been carried out with several radio telescopes, including the (Karl G.~Jansky) VLA (e.g. Bondi et al.,~2003; Simpson et al., 2006; Schinnerer et al., 2007; Miller et al., 2013; Heywood et al., 2013), the Westerbork Synthesis Radio Telescope (WSRT; e.g. de Vries et al., 2002) the Giant Metrewave Radio Telescope (GMRT; e.g. Garn et al., 2007) and the Australia Telescope Compact Array (ATCA; e.g. Norris et al., 2006; Middelberg et al., 2008).

The Square Kilometre Array (SKA; Dewdney et al., 2013) promises to revolutionise our understanding of star formation (Jarvis et al., 2015a) and AGN processes (Smol{\v c}i{\'c} et al., 2015) across cosmic time, as well as truly realise the potential that deep \emph{and} wide radio continuum surveys have for answering key questions in cosmology (Jarvis et al., 2015b). As we move towards construction of the SKA, a new generation of large scale radio continuum surveys are being planned and executed with new SKA pathfinder instruments, as well as through significant hardware upgrades of some existing radio telescopes. The increased capabilities of these machines over their predecessors, -- typically some combination of an expanded field of view, more sensitive receivers and a huge increase in instantaneous bandwidth -- will allow surveys with depth or areal coverage that improve on existing observations by orders of magnitude.

At the deep end the MIGHTEE survey on the MeerKAT telescope (Booth \& Jonas, 2012) aims to cover 35 square degrees to a depth of 1~$\mu$Jy beam$^{-1}$ in its deepest tier (Jarvis, 2012). Complementary to such deep observations are the all-sky surveys: the ASKAP (Johnston et al., 2008; de Boer et al., 2009) EMU survey (Norris et al., 2011) aims to cover the entire sky south of declination +30$^{\circ}$ to a depth of 10$\mu$Jy with 10$''$ angular resolution. The WODAN survey (R{\"o}ttgering et al., 2011) will use APERTIF (Oosterloo et al., 2009; van Cappellen \& Bakker, 2010), a hardware upgrade to the WSRT, to complete the full sky coverage by conducting a similar survey in the northern hemisphere. The successor to NVSS is also being planned for the VLA\footnote{{\tt https://science.nrao.edu/science/surveys/vlass}}, and at the time of writing is envisaged to consist of an all-sky snapshot survey at 2--4~GHz, reaching a depth of 69~$\mu$Jy beam$^{-1}$, and taking advantage of the extended configurations of the VLA to reach an angular resolution of 2.5$''$ (Condon, 2015). These will be complemented by large-area surveys at low radio frequencies using aperture arrays, including the now largely complete 30--160~MHz MSSS survey using the Low Frequency Array (Heald et al., 2014) and the GLEAM survey with the Murchison Widefield Array (Wayth et al., 2015) at 80--230~MHz.

The ASKAP and APERTIF telescopes both feature phased array feed (PAF) receivers, in which an array of many receptors is placed in the focal plane of each of the dishes. The voltages from these multi-element receptors are linearly combined with multiple sets of complex weights and summed to generate multiple beams, steered in different directions within the field of view of the instrument. PAF beams from each dish are cross-correlated with those sharing the same direction from all other dishes. This parallel processing of multiple beams results in a dramatic increase in field of view (and therefore survey speed) over an equivalent single pixel feed instrument. 

In this paper we present the results of a pilot, three epoch, broadband (711--1015 MHz) continuum imaging survey covering approximately 150 square degrees in the constellation of Tucana, and encompassing about two-thirds of the \emph{Spitzer} South Pole Telescope Deep Field (Ashby et al., 2013), using the Boolardy Engineering Test Array (BETA). This is a prototype of the ASKAP array, consisting of 6 of the 36 dishes, equipped with the first generation (Mark I) PAF system (Schinckel et al., 2011) based on a connected-element `chequerboard' array (Hay \& O'Sullivan, 2008). A detailed description of BETA is provided by Hotan et al.~(2014). 

We describe the observations in Section \ref{sec:obs} and the calibration and imaging procedure in Section \ref{sec:reduction}. The data products are described in Section \ref{sec:results}. The photometric, astrometric and spectroscopic performance of BETA is examined in detail in Sections \ref{sec:photometry}, \ref{sec:astrometry} and \ref{sec:spectra}. For these purposes we primarily make use of SUMSS, carried on using the Molonglo Observatory Synthesis Telescope (MOST) which is well matched to the BETA observations in terms of frequency and angular resolution. We also make use of lower frequency observations with the GMRT and higher frequency observations from the VLA, chiefly NVSS and FIRST. Concluding remarks are made in Section \ref{sec:conclusion}.

\section{Observations}
\label{sec:obs}

\begin{figure}
\centering
\includegraphics[width=\columnwidth]{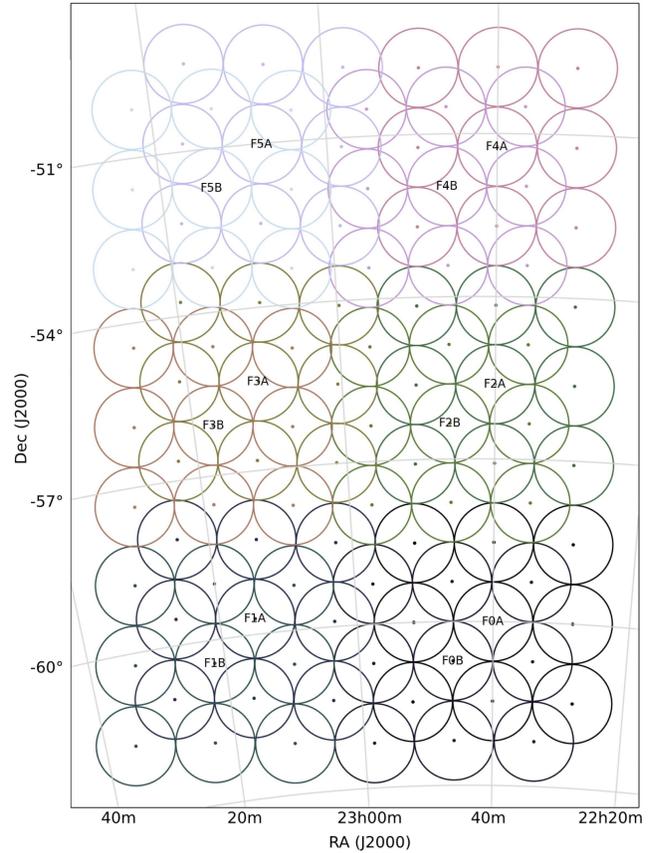}
\caption{The sky area covered by the observations is shown above. Coverage is achieved through a combination of twelve array pointing centres (as labelled) and the nine simultaneous beams associated with each of them, the circles showing the approximate half power point of the beams at the band centre (863~MHz). The nine beams are placed in a 3~$\times$~3 square arrangement, and those associated with each pointing centre are represented by a common colour on this plot.}
\label{fig:ptgs}
\end{figure}

The target field was observed with BETA on three separate occasions as part of the commissioning and verification of the instrument. The telescope delivers 304 MHz of instantaneous bandwidth and for these observations the sky frequency range was 711--1015 MHz, corresponding to a fractional bandwidth of 35\%. The data are captured with a frequency resolution of 18.5~kHz, using 16,416 frequency channels across the band.

The digital beamformers of BETA (Bunton et al., 2011; Hampson et al., 2011) are capable of delivering nine dual-polarisation beams that can be placed arbitrarily within the $\sim$30 square degree field of view of the instrument. At present a maximum signal-to-noise algorithm is employed to form the compound beams (Applebaum, 1976; Hotan et al., 2014). In brief, this approach requires the PAF elements to be excited to high significance by a strong signal. The Sun is appropriate for such a purpose on a 12~m dish. The direction of a given beam is enforced by steering the antenna so that the Sun lies along that direction, and determining complex weights for each of the 188 PAF elements (94 in each orthogonal mode of linear polarisation) to maximise signal from the Sun with respect to the system noise. In the case of these observations a regular 3~$\times$~3 footprint of beams was employed, with a central on-axis beam surrounded by eight additional beams on a square grid with a spacing of 1.46 degrees between the centre of each beam.

The spacing of the grid was chosen to be approximately the half-power beam width of a single PAF beam at the band centre. For `traditional' mosaicking of a region of sky using a single pixel feed interferometer, the array will typically observe a list of discrete positions that have the primary beam from one pointing situated at the half power point of the adjacent scan, typically with a hexagonal arrangement. In the case of ASKAP, the three-axis mount on the antennas in the array keeps the deployed beam pattern fixed on the sky relative to the antenna pointing direction (unless the weights are adjusted). Thus an appropriate combination of beam pattern and pointing positions can be devised to rapidly cover a large area of sky to approximately uniform depth. 

For this project the BETA array spent five minutes on each of twelve sky positions, repeating the cycle for the 12 hour duration of the observation. The pointing positions are arranged in six close pairs, with the close pairs used to offset the fact that the beam spacing at any given pointing is twice the value that would generally be used for a standard mosaic (Bunton \& Hay, 2011; Hay \& Bird, 2015). The end result is the approximately uniform sky coverage shown in Figure \ref{fig:ptgs}, effectively mimicking a traditional 108 pointing radio survey using only twelve BETA pointings. Groups of nine beams share the same colour in this plot.

\begin{figure}
\centering
\includegraphics[width=\columnwidth]{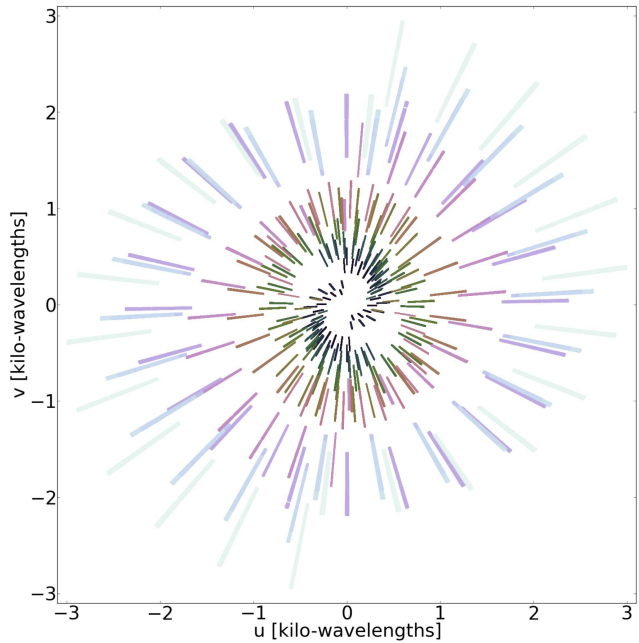}
\caption{The ($u$,$v$) plane coverage of a typical pointing, in this case field F0A from Scheduling Block 1231. Note that this ($u$,$v$) plane coverage is shared by all nine beams associated with that pointing. Interleaving each of the twelve pointings via the constant rotation of five minute scans builds up good Fourier plane coverage. The radial coverage is due to the 304~MHz of bandwidth. Each baseline has a unique colour on this plot.}
\label{fig:uvplot}
\end{figure}

Cycling around the twelve pointing centres with short five minute integrations over a twelve hour observation builds up favourable Fourier plane coverage for each of the pointings, which is shared by the nine beams associated with that pointing. The ($u$,$v$) plane coverage of a single pointing is shown in Figure \ref{fig:uvplot}. The radial coverage afforded by the 304~MHz of bandwidth is apparent, and the points are coloured per baseline. Note that the shortest baseline has been removed for reasons explained in Section \ref{sec:preprocessing}.

The ASKAP antennas are equipped with an additional axis of movement (the roll axis; Forsyth et al., 2009) that keeps the parallactic angle fixed over the course of an observation, and thus keeps the beam pattern fixed on the sky without the need for continual adjustment of the weights (Hay, 2011). For each of the pointing positions for this survey the roll axis position was adjusted to remove the parallactic angle offset from scans that occur along lines of fixed declination. The end result is a more regular survey area that does not taper with declination.

The survey area was observed with BETA on three separate occasions. Table \ref{tab:epochs} lists the start and end times of the ASKAP Scheduling Blocks (SBs) used in the project. In addition to the observations of the target field a calibration scan is performed whereby the array executes a pointing pattern that places the standard calibrator source PKS B1934$-$638 at the nominal centre of each beam for 5 minutes. These scans (marked with the asterisks in Table \ref{tab:epochs}) are used to calibrate the bandpass response of each beam and to set the flux density scale.\footnote{This approach would likely prove too costly to be executed on a per-observation basis for the full 36-beam ASKAP array. The calibration of the beams for full ASKAP is likely to make use of incremental corrections to the solutions derived from an initial reference calibration observation of a strong calibrator source. The incremental solutions will be derived by making use of the on-dish calibration system, an artificial radiator at the centre of the primary reflector that is used to illuminate the PAF elements to high significance. This feature is not part of the BETA hardware.} Note that SBs 1206 and 1207 were merged into a single data set. With a total duration of 8h 38m this observation is shorter than those of SBs 1229 and 1231. Further details of the calibration process are provided in Section \ref{sec:reduction}. The beamforming process was initiated at 05:00 UT on 2 December 2014, immediately prior to the first astronomical observation performed as part of this pilot survey, as listed in Table \ref{tab:epochs}. No further updates to the beamformer weights were made during these observations.

\begin{table}
\centering
\caption{Start and end times and dates (UT) for the ASKAP Scheduling Blocks (SBs) that were used in this project. SBs marked with an asterisk (*) indicate that the observations consist of per-beam scans of the standard calibrator source PKS B1934$-$638, as described in the text. Note that SBs 1206 and 1207 are part of the same observing run which interrupted for approximately an hour. In total this first epoch has 8h 38m of data making it shallower than the subsequent two epochs.}
\begin{tabular}{llllll} \hline
SB    & Date        & Start (UT) & End (UT) & Duration (h) \\ \hline
1205* & 02-Dec-2014 & 06:25:39   & 07:12:39 & 0.78  \\
1206  & 02-Dec-2014 & 07:14:44   & 09:56.49 & 2.70  \\
1207  & 02-Dec-2014 & 11:03.04   & 16:59:44 & 5.94  \\ 	
1227* & 07-Dec-2014 & 03:21:01   & 04:07:56 & 0.78  \\
1229  & 07-Dec-2014 & 04:16:06   & 16:30:46 & 12.24 \\
1230* & 08-Dec-2014 & 04:09:46   & 04:56:41 & 0.78  \\
1231  & 08-Dec-2014 & 05:00:41   & 17:04:51 & 12.07 \\ \hline
\end{tabular}
\label{tab:epochs}
\end{table}

\section{Data reduction}
\label{sec:reduction}

The procedure for calibrating and imaging the data was identical for each of the three epochs, and essentially followed standard practice for modern radio interferometers, albeit with a few special considerations due to the multiple primary beams that BETA delivers. It was achieved with a custom software pipeline that uses components from numerous packages, as described in this section. The procedure was entirely automatic. Note that the steps described in Sections \ref{sec:selfcal} and \ref{sec:imaging} were executed twice. Following the first run an automatic source finder was used, the positions of the detected sources were turned into a mask in order to constrain the deconvolution. Once the cleaning masks were in hand, the initial calibration was discarded and the process restarted, with the cleaning masks employed throughout the second run. The necessity of this step is discussed in Section \ref{sec:biases}. 

The bespoke calibration and imaging procedure used for the BETA continuum data is the result of testing numerous approaches. The methods outlined below are not expected to be those adopted for the full ASKAP array. ASKAP has a dedicated software system (ASKAPsoft) that is a crucial component of the telescope, designed specifically to continuously process the data streams in pseudo real time. ASKAPsoft is itself undergoing commissioning tests, and in any case it relies on many properties of the final array (e.g.~low point spread function (PSF) sidelobe levels) that cannot be met by BETA. A complete description of the ASKAP data processing model is provided by Cornwell et al.~(2011).

\subsection{Pre-processing}
\label{sec:preprocessing}

The BETA telescope produces approximately 2~TB of visibility data per 24 hours of observing. Each Scheduling Block produces a single {\tt CASA} format Measurement Set that contains cross and autocorrelation measurements for all nine beams, with distinctions between beams encoded via the {\tt FEED} table. All four linear polarisation products are produced (XX, XY, YX and YY). The visibilites for each beam were split into a standalone Measurement Set using the {\tt CASA}\footnote{{\tt http://casa.nrao.edu}} package (McMullin et al., 2007). Initial flagging operations were also applied at this stage using the {\tt flagdata} task, namely the removal of the autocorrelations, the deletion of the shortest baseline\footnote{Antennas AK01 and AK03 are separated by only 37~m. Including this baseline tends to degrade the quality of the image due to the introduction of emission on large spatial scales that is not faithfully reproduced due to the lack of spacings between the shortest two baselines of the BETA array. This situation will improve as more antennas in the core of ASKAP come online.}, amplitude thresholding and a single pass of the {\tt rflag} algorithm\footnote{{\tt rflag} identifies and excises errant visibility points based on their deviation from statistics computed in sliding time and frequency windows.} as implemented in {\tt CASA}.

\subsection{Bandpass and flux scale calibration}
\label{sec:fluxcal}

The per-beam scans of PKS B1934$-$638 at full spectral resolution were averaged in time, and per-channel complex gain solutions were derived that best correct the observed data to the model polynomial fit to the spectrum of the source derived by Reynolds (1994). Two passes of a sliding median filter were applied to the derived corrections to remove errant channels. The frequency behaviour of these solutions thus corrects the effective bandpass of each of the formed beams, and since the solutions were not normalized they also encompass the flux density scaling of the target data. The per-beam observations of the target field were corrected by applying the relevant calibration table derived from PKS B1934$-$638. Motivated by the large field of view afforded by the 12~m dish at these observing frequencies, we have determined via simulations that the $\sim$Jy-level sources in the field of PKS B1934$-$638 manifest themselves as a noise-like signal at the $\sim$1\% level when averaged over 5 minutes of data, irrespective of the hour angle, and do not need to be included in the calibration model.

\subsection{Self-calibration}
\label{sec:selfcal}

Following bandpass and flux scale correction the target data were averaged from 16,416~$\times$~18.5~kHz channels to 304~$\times$~1 MHz channels. No time averaging was performed, preserving the default correlator integration time of 5 seconds. At this stage further improvements to the data quality must be made via self-calibration techniques as no additional calibration scans were made. This proceeds as follows, for each of the twelve fields, each of which has nine corresponding beams, giving a total of 108 datasets per target SB that must be independently calibrated. 

The {\tt CASA} {\tt clean} task was used to make a multi-term, multi-frequency synthesis (MT-MFS; Rau \& Cornwell, 2011) image of the data in four spectral chunks, each having 76~MHz of bandwidth: 711--787, 787--863, 863--939, 939--1015 MHz respectively for sub-bands 1, 2, 3 and 4. As mentioned in Section \ref{sec:obs}, the roll axis of the ASKAP telescope keeps the beam pattern fixed relative to the sky. Having sidelobes that do not rotate offers considerable advantages in terms of using standard direction-independent self-calibration techniques to deal with sources in the sidelobes  (e.g.~Smirnov, 2011; Heywood et al., 2013). However for a typical BETA observation in this band substantial effective depth is achieved through the first sidelobe and numerous sources are detectable. The image must therefore be wide enough to allow these sources to be deconvolved. The $w$-term is corrected for during gridding (Cornwell, Golap \& Bhatnagar, 2005) to more faithfully recover far field sources. Also, since the purpose of these initial images is to form a sky model for refining the calibration, far field sources must also be characterised to that end.

For a single SB this produces 12 fields $\times$ 9 beams $\times$ 4 sub-bands = 432 images. A sky model was constructed by using the {\tt PyBDSM} (Python Blob Detection and Source Measurement; Mohan \& Rafferty, 2015) source finder to decompose the Stokes I images into a series of point and Gaussian components. The source finder first estimates the spatial variation in the RMS noise ($\sigma$) in the image by stepping a box across the image, measuring the standard deviation of the pixels within it and then interpolating the measurements. This distribution is then subsequently used by the source finder for thresholding purposes. Image peaks that exceed a user-specified threshold (in this case 7$\sigma$, where $\sigma$ is the local RMS value) are identified. These peaks are then grown into islands, contiguous regions where the pixel values exceed a secondary threshold (in this case 5$\sigma$). Having identified these islands {\tt PyBDSM} attempts to fit them with point and Gaussian components, and catalogue and image products describing the fit can be exported in various formats. Manual checks when designing and refining the calibration and image pipeline verified that the thresholds used returned a largely complete model that did not include spurious features. No assumptions about the primary beams of the telescope have been made at this stage, thus the model captures the apparent rather than the intrinsic brightnesses of the sources in the field.

The {\tt MeqTrees} package (Noordam \& Smirnov, 2010) was used to predict model visibilities based on the per sub-band sky models derived from the image by {\tt PyBDSM}. A set of per-antenna phase-only complex gain corrections were then solved for by comparing the observed visibilities to the model data. An independent correction was derived for both the XX and YY polarisations, although no polarisation information was included in the model. A single solution was derived for each five minute scan and for each sub-band.

\subsection{Imaging}
\label{sec:imaging}

Following the correction of the data using the self-calibration solutions each of the 108 individual Measurement Sets were re-imaged and deconvolved in four spectral sub-bands using the {\tt CASA clean} task in a similar way to the process that was used to derive the sky model. For the final images deconvolution was performed in two passes, a deep\footnote{For the initial deep deconvolution the clean cycle is terminated when the peak residual is below 5 times the RMS noise of the corresponding Stokes V image (see Section \ref{sec:sensitivity}), or at 5,000 iterations, whichever occurs first.} clean with the mask in place and a secondary shallow\footnote{The shallow cleaning operation consists only of 100 blind iterations on the residuals remaining following the initial deep clean.} pass with the mask removed. A common Gaussian restoring beam of 70$''$~$\times$~60$''$ (PA~=~0$^{\circ}$) was enforced for all images.

Primary beam correction was achieved using the standard Airy pattern for a 12~m dish appropriate for each sub-band, as computed by the {\tt CASA} imager (see also Appendix A). One quirk of the BETA system\footnote{This will not be the case for any ASKAP array that is equipped with Mark II PAFs.} is that each beam records visibilities that are fringe stopped to a common reference direction, typically that of the array pointing centre, which in this case is coincident with the centre of the on-axis formed beam. Thus when imaging off-axis beams with a standard imaging package the region of maximal sensitivity will be offset from the image centre at the nominal centre of the beam that has been imaged. The primary beam images must be shifted to the correct position before the correctional division operation occurs. Corrected images were cut at the radius where the primary beam gain dropped below 30\%. Linear mosaicking was done with the {\tt Montage}\footnote{{\tt http://montage.ipac.caltech.edu/}} software, using the assumed variance patterns as weighting functions. 

\subsection{A note on calibration and deconvolution biases}
\label{sec:biases}

With the removal of the shortest baseline, self-calibration of each BETA beam involves solving for six unknowns (the per-beam complex gain terms) using only 14 equations (one per baseline, e.g.~Cornwell \& Wilkinson, 1981; Ekers, 1984). The self-calibration problem for BETA is therefore relatively poorly constrained when compared to an array such as the VLA (27 antennas, 351 baselines), ASKAP (36 antennas, 630 baselines) or MeerKAT (64 antennas, 2016 baselines). A careful and conservative approach must be adopted in order to avoid self-calibration biases, whereby the contribution to the visibility measurement made by sources that are not in the sky model can be subsumed into the gain corrections causing that set of (typically fainter) sources to be suppressed. Incomplete sky models can also impart subtle and highly non-intuitive features into a radio image, see for example Grobler et al.~(2014).

Furthermore the phenomenon of clean bias (or snapshot bias) must also be considered. This also typically manifests itself as a systematic underestimation of the source flux densities above some threshold, and has been shown to exhibit flux-dependent suppression of the fainter sources, as well as affect the flux densities of sources below the thermal noise limit (White et al., 2007). Most investigations into clean bias have been conducted using VLA data (where strong linear features in the snapshot PSF are thought to exacerbate the problem) via the injection of synthetic sources with known brightness into the data (Condon et al., 1998; Becker, White \& Helfand, 1995). A thorough understanding of the problem in the context of broadband radio continuum imaging is yet to be achieved, however Helfand, White \& Becker~(2015) note that bias appears to correlate with interference levels.

Blind cleaning of the BETA data with varying number of iterations shows that the recovered flux densities of sources (particularly at the faint end) can be strongly affected by the deconvolution, likely due to the high ($\sim$15\% of peak) PSF sidelobe levels of the six dish array. Thus we only apply shallow deconvolution to the residual data following the deconvolution step that employed masks. This results in accurate flux density measurements (Section \ref{sec:photometry}) and reliable automatic pipelining of the data, however the price we pay for this is an effectively higher noise floor (Section \ref{sec:sensitivity}).

The symptoms introduced by these biases are temporary, in that they are expected to be greatly reduced as more antennas in the ASKAP array come online. The self-calibration procedure will be better constrained and the deconvolution process will be much more robust as the PSF side lobes are significantly reduced due to the improved ($u$,$v$) plane coverage. It is pleasing to note that the most troublesome aspects of processing the data from this new telescope occurred simply because we only had six antennas at our disposal. 

\section{Results and discussion}
\label{sec:results}

\subsection{Wide field radio images}

\begin{figure*}
\begin{center}
\setlength{\unitlength}{1cm}
\begin{picture}(16,22.4)
\put(-0.84,-0.76){\includegraphics{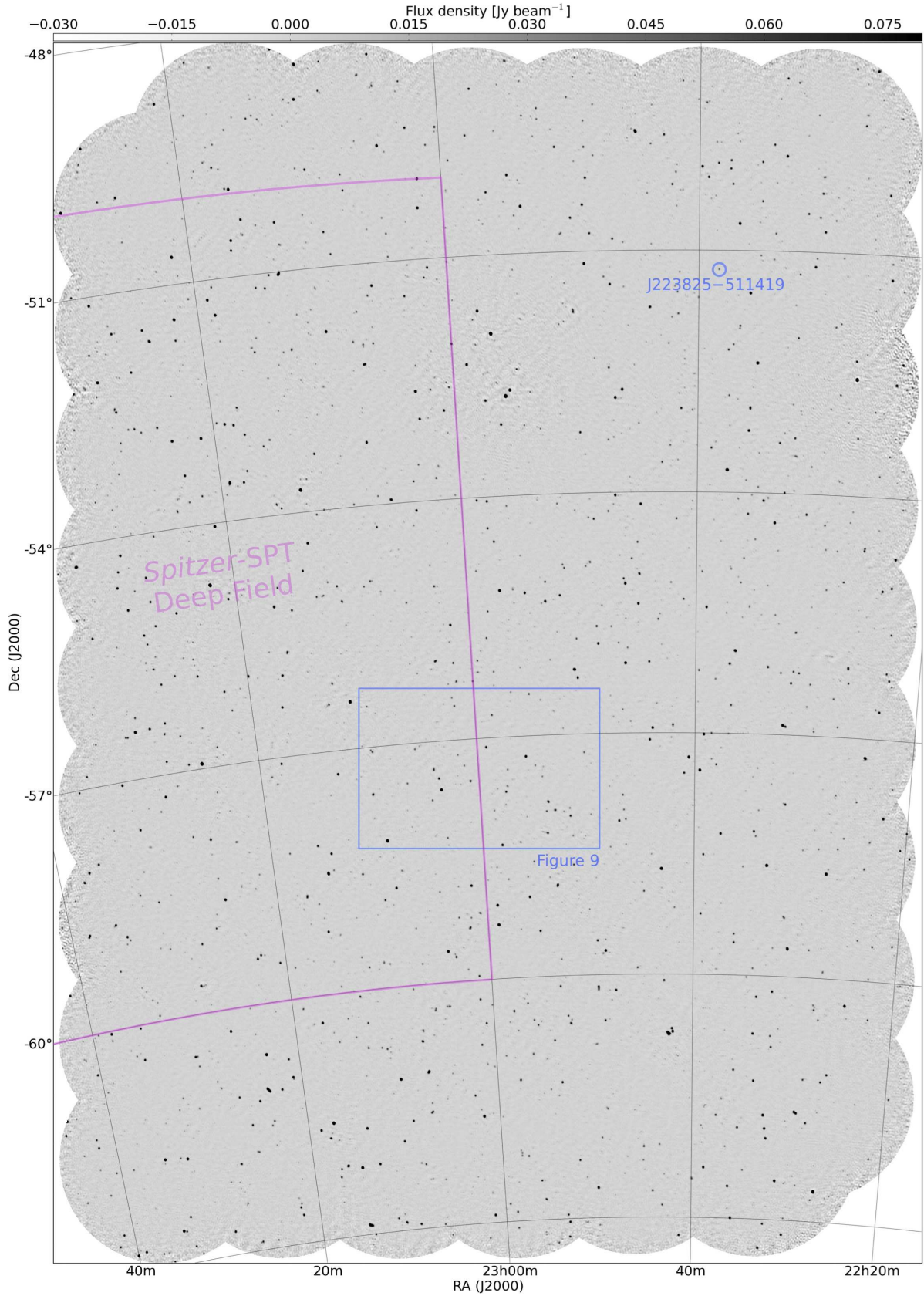}}
\end{picture}
\caption{Total intensity mosaic formed from combining the images from all three epochs. The rectangle marks the area presented in Figure \ref{fig:zoom} and Section \ref{sec:spectra}, and the area of overlap with the \emph{Spitzer} South Pole Telescope Deep Field is shown. The single significantly variable source J223825$-$511419 (Section \ref{sec:variability}) is also marked. The greyscale is linear and runs from $-$30 to 80 mJy beam$^{-1}$.}
\label{fig:mosaic}  
\end{center}
\end{figure*}

The automated calibration and imaging procedure described in Section \ref{sec:reduction} was applied to each of the three observations of this field, resulting in 432 primary beam corrected images per epoch. Four sub-band mosaics were produced from the 4~$\times$~76 MHz chunks of calibrated data across the 304~MHz band: . The full band mosaics were formed by linearly mosaicking all 432 images per epoch in order to include a (somewhat coarse) frequency-dependent primary beam correction. These full band mosaics form the basis of the instrumental verification presented in Sections \ref{sec:photometry} and \ref{sec:astrometry}, as well as the variability study presented in Section \ref{sec:variability}. A combined, deep image was generated by stacking the mosaics formed from the three epochs. This is used to produce a catalogue of the components in the image as described in Section \ref{sec:catalogue}, as well as to measure the differential source counts at 863 MHz (Section \ref{sec:sourcecounts}). Combined three-epoch sub-band mosaics were also produced. These were used to examine the in-band spectral performance of BETA in Section \ref{sec:spectra}, as well as produce estimates of the source spectral indices for the catalogue. Prior to combining the images, astrometric corrections were applied in order to correct minor offsets in the coordinate frames, as described in Section \ref{sec:astrometry}. 

To summarise: each of the three epochs resulted in five mosaics (four sub-band and one full band) and there are an additional five mosaics for the combined data, twenty images in total. The combined epoch, full band radio mosaic is shown in Figure \ref{fig:mosaic}. The {\tt PyBDSM} source finder using a pixel threshold of 5$\sigma$ and an island threshold of 3$\sigma$ (where $\sigma$ is the source finder's own estimate of the local noise, see Section \ref{sec:selfcal}) was used to produce component lists for each of these images for cross-matching and verification purposes.

\subsection{Sensitivity and confusion}
\label{sec:sensitivity}

Estimates of the local background noise produced by {\tt PyBDSM} serve to deliver a source catalogue that has very high reliability, and the resulting RMS maps are very useful for computing the visibility areas at different flux thresholds (Section \ref{sec:sourcecounts}). They may not however be a reliable indicator of the thermal noise performance of the observation, and more accurately quantify the `effective' noise of the image. In the case of the BETA observations the effective noise limit is dominated by three effects. The first is typical of radio interferometer maps in that the regions around bright sources tend to have elevated artefact levels due to calibration deficiencies. The other two effects at play here are deconvolution related: the conservative approach to calibration and deconvolution employed in order to minimise biases (Section \ref{sec:biases}) results in only shallow cleaning of the faint sources in the image. Residual sidelobes thus contribute significantly to the image background. The second effect comes about by our use of image-plane combination of the data across the band. As the true noise floor is pushed down, fainter sources are revealed that have not been cleaned at all.

Classical source confusion is not expected to be contributing to the effective noise in these images. The level of classical confusion can be estimated by using the extragalactic radio continuum simulation of Wilman et al.~(2008; 2010). The simulation uses observed and extrapolated luminosity functions to generate mock galaxy populations including radio loud and quiet AGN, quiescent star forming galaxies and Gigahertz Peaked Spectrum (GPS) sources, the clustering properties of which are determined by a model of the underlying dark matter distribution. The result is a catalogue of $\sim$260 million components over 400 square degrees with a flux limit of 10~nJy, with each component having an estimate of its radio flux density at five frequencies.

We linearly interpolate the simulated 610 and 1400 MHz flux density measurements to 863 MHz, and determine the flux limit at which the number of sources per unit area exceeds a threshold at which the observations are deemed to be confused. From this we estimate that the classical confusion limit for the BETA observations is 337 $\mu$Jy for $m$~=~10, and 180 $\mu$Jy for $m$~=~5, where $m$ is the number of beams per source, the criterion typically used to define classical confusion (Wilson, Rohlfs \& H\"uttemeister, 2013). Simulations based on the faint source count measurements of Condon et al.~(2012) estimate the classical confusion limit to be 158 and 202~$\mu$Jy~beam$^{-1}$ at 711 and 1015~MHz respectively for a resolution of 70$''$~$\times$~60$''$ (J. Condon, private communication).

We can estimate the true thermal noise performance of the data by forming Stokes V images and reproducing the mosaics. More accurately these maps will be pseudo-Stokes-V, as although use of the unpolarised calibrator PKS B1934-638 leads to accurate calibration of the XX and YY gains, no further polarisation calibration has taken place, and some instrumental leakage from other Stokes parameters into V will be present. Thus even intrinsically unpolarised sources will leave some fraction of their Stokes I flux in the Stokes V image, the level of which is coupled to the difference between the true XX and YY beams and the assumed primary beam used to correct the images at the position of the source. The error introduced by these differences typically increases with distance from the centre of the beam, thus spurious emission in the Stokes V mosaics is effectively suppressed by the mosaic weighting scheme. Sources that are both bright and strongly intrinsically polarised will also in principle remain in the image, typically at a level that is a few percent of their Stokes I brightness.

\begin{figure}
\centering
\includegraphics[width=\columnwidth]{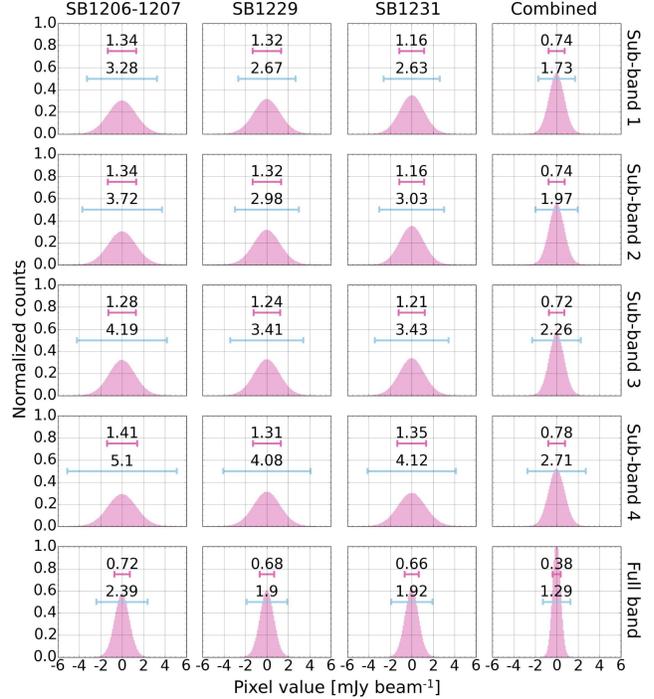}
\caption{The pink histograms of the pixel intensities of the pseudo-Stokes-V mosaics provide effective measurements of the thermal noise of the observations, the $\pm$1$\sigma$ values of which are indicated by the upper bars on each subplot. The effective noise level (estimated by {\tt PyBDSM} from the Stokes I mosaics) is elevated by a factor of $\sim$3, primarily due to incomplete deconvolution, as indicated by the lower bars on each subplot. The columns are the three epochs and the combined data, and the rows show the four sub-bands plus the full-band data. Refer to the text for full details.}
\label{fig:rmsplots}
\end{figure}

Figure \ref{fig:rmsplots} shows histograms of pixel intensity as measured from the pseudo-Stokes-V mosaics over regions that have more than one beam contributing to them. The columns are the three epochs and the combined data, and the rows show the four sub-bands plus and the full-band data. The sensitivity gain afforded by averaging in time (first three columns into the final column) or frequency (first four rows into the final row) is clear. The labeled pink bars show the $\pm$1$\sigma$ values, where $\sigma$ is the standard deviation of the pixel histograms. We assume that these measurements give an accurate measurement of the thermal noise (from the effective absence of any astronomical signal). Also shown on the plot via the blue bars are the median values of the noise measurements from the corresponding {\tt PyBDSM} Stokes I RMS map. The effective noise is typically a factor of $\sim$3 higher than the expected thermal noise.

This is undesirable, but is a penalty that is paid in exchange for the autonomous pipelining of the data. Human-guided calibration and imaging of BETA data can successfully deal with this issue, primarily by supervised deconvolution and identification of spurious image features, and the manual construction of deep sky models. We note that sidelobe confusion, as with the calibration and deconvolution biases, will be much reduced in future versions of the ASKAP array.

\subsection{Photometry}
\label{sec:photometry}

\begin{figure}
\centering
\includegraphics[width=\columnwidth]{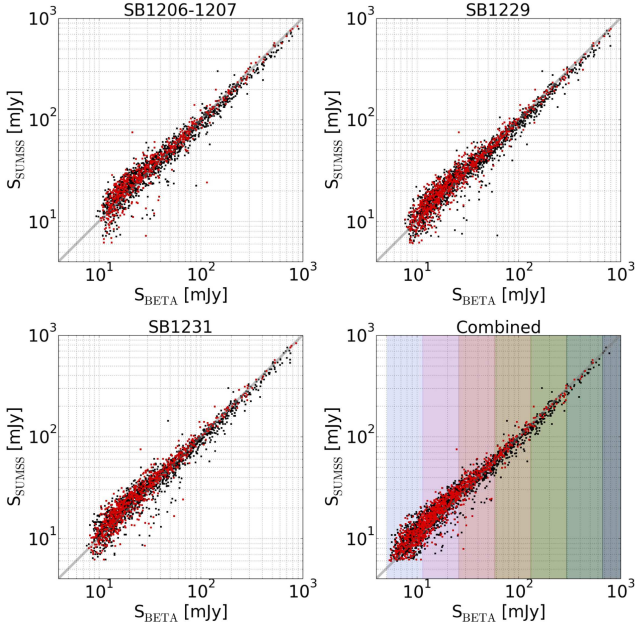}
\caption{Peak flux densities of the BETA components compared to those of matched components from the SUMSS catalogue for the three epochs and the combined data. The red points indicate components that appear unresolved in SUMSS (and should therefore be unresolved in the BETA data). The hard lower edge on the combined data plot (lower right) is due to the 6~mJy beam$^{-1}$ peak flux limit of SUMSS. The number of matched components for the four panels are (left to right, top to bottom) 1,841; 2,282; 2,241 and 2,927. The relative shallowness of the first epoch and the increased depth of the combined data are reflected in these counts. The coloured bars on the combined panel denote the ranges of the flux density bins used to construct Figure \ref{fig:ratios}.}
\label{fig:sumss_comp}
\end{figure}

\begin{figure}
\centering
\includegraphics[width=\columnwidth]{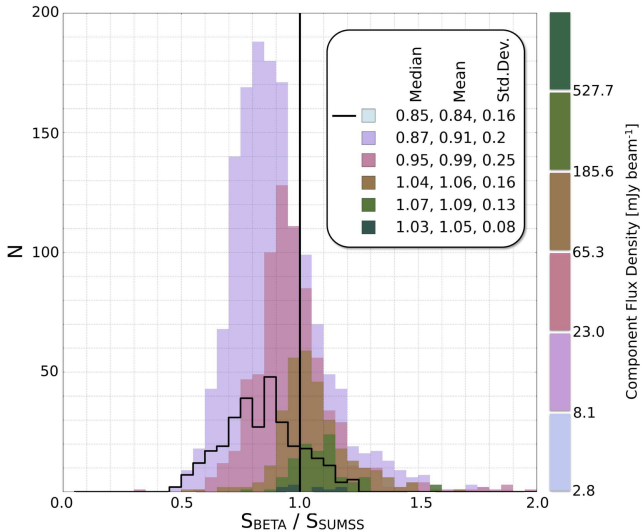}
\caption{Histogram of component counts as a function of their BETA to SUMSS peak flux density measurements. The sources are grouped into logarithmically-spaced bins according to their BETA peak flux density, as indicated by the key above and by the stripes on the lower right panel of Figure \ref{fig:sumss_comp}. The faintest bin is represented by the black line histogram for clarity. Values on the figure show the median, the mean and the standard deviation of the ratios per bin.}
\label{fig:ratios}
\end{figure}

As with a traditional single pixel feed interferometer, regular observations of a calibrator source are required to determine the absolute flux density scale in the presence of temporal instrumental gain drifts. With a PAF system however the primary beams are formed from weighted sums of the signals from many individual elements. While the directional responses of the individual elements can be assumed to be stable (as can to first order the primary beam response of a single pixel feed telescope), the temporal drifts in the electronic gains of these elements have the potential to modulate the shape of the compound beams (Smirnov \& Ivashina, 2011). While the absolute flux density response of the array can be corrected by visiting a known calibrator, correcting variations in the off-axis response due to element gain drifts can only be done by making compensatory adjustments of the beamformer weights\footnote{Such a system will be implemented for ASKAP's Mark II PAF system through the use of the on-dish calibration system, allowing a noise source to be radiated onto the PAF from the centre of the primary reflector. Direction dependent calibration schemes are another viable method for dealing with these effects.}. For the survey presented in this paper, the BETA beamformers were loaded with appropriate weights prior to the first observation and these weights were not adjusted for the week over which the observations were conducted. 

We examine the photometric accuracy and stability of our observations by cross matching the list of sources detected in the full band, per-epoch radio mosaics, as well as in the final combined mosaic, with the SUMSS catalogue. The 843 MHz observing frequency of SUMSS is slightly offset from our 863 MHz observing frequency, however the fractional change in the flux density of a typical $\alpha$~=~$-$0.7 ($S~\propto~\nu^{\alpha}$) source between the observing frequencies of SUMSS and these observations is less than 2\%. We do not attempt to correct for this. Components were matched by searching for the pairwise minimum separation, and requiring that this separation be less than 20$''$, or roughly one third of the BETA synthesised beam width. In addition, contiguous islands of emission in the BETA catalogues that contained more than a single fitted component were rejected. This has the effect of only selecting isolated sources for the cross matching, avoiding regions of emission that may be fitted by different multiple components between the surveys, although with the somewhat low angular resolution of these observations such complex sources are rare.

Figure \ref{fig:sumss_comp} shows a log-log plot of the peak flux densities of the SUMSS components against those of the matching BETA components for the three epochs, as well as the combined epoch as marked above each panel. The diagonal line is the 1:1 relationship. Increased scatter in the measurements with decreasing flux density is to be expected as the noise of the observations becomes an increasingly significant fraction of the component brightness. In addition to this feature, the distributions in Figure \ref{fig:sumss_comp} show a non-linear shift from an excess in the BETA measurements at the bright end to SUMSS excess at the faint end. This shift is visualised and quantified in Figure \ref{fig:ratios} which shows histograms of the component counts as a function of their BETA to SUMSS peak flux density ratios. These are grouped into logarithmically-spaced bins according to their BETA peak flux density measurements, as indicated on Figure \ref{fig:ratios}, and by the coloured bars in the lower right panel of Figure \ref{fig:sumss_comp}. The numbers on Figure \ref{fig:ratios} show the median, mean and standard deviation of the ratios for each flux density bin, moving from a few percent excess in the BETA measurements at the bright end to a $\sim$15\% decrement in the faintest bin. 

Such shifts away from the 1:1 relationship are often seen when comparing flux density measurements from surveys with mismatched angular resolution and depth. For example, Franzen et al.~(2015) compare ATCA flux density measurements to deeper VLA data for a sample of unresolved sources, noting an excess in the ATCA measurements that approaches a factor of 1.5 times the VLA flux density for the faintest sources. This is ascribed in part to Eddington bias (Eddington, 1913) skewing the measurements high in the shallower ATCA data.
Allison, Sadler \& Meekin (2014) compare SUMSS data to matched sources from the single dish Continuum HI Parkes All Sky Survey catalogue (Calabretta, Staveley-Smith \& Barnes, 2014), resulting in a distribution that is qualitatively very similar to those in Figure \ref{fig:sumss_comp}, and Hodge et al. (2011) see a trend towards excess brightness in their VLA A-array measurements compared to the lower resolution FIRST measurements which they interpret in terms of resolution biases. 

Due to projection effects, the resolution of SUMSS is 45$''$~$\times$~(45 csc $|\delta|$)$''$ where $\delta$ is the declination, increasing the north-south size of the PSF from 50$''$ to 60$''$ over the extent of the common area. However for all positions the PSF of the SUMSS survey is (a factor of 1.1--1.5 times) smaller than that of BETA. SUMSS resolves $\sim$10\% of sources ($\sim$25\% at its southernmost declinations, $\sim$2\% towards the celestial equator). No clear trend emerges in examinations of the difference between the BETA and SUMSS measurements as a function of the deconvolved source major axis. Selecting sources that are unresolved in SUMSS by requiring a peak to integrated flux ratio of between 0.98 and 1.02 results in the subset of objects denoted by the red points on Figure \ref{fig:sumss_comp}, and a two-sample Kolmogorov-Smirnov test rejects the hypothesis that these are drawn from distinct distributions ($p$-value = 15\%). If resolution bias is at play here it is not likely to be the dominant cause of the observed trend.

Serra et al.~(2015) used 30 hours of BETA time to conduct HI imaging of a nearby galaxy group over the frequency range 1.4025--1.4210~MHz (1.3\% fractional bandwidth). Cross referencing of the sources detected in the corresponding continuum image with the NVSS catalogue revealed a median and mean excess of 7\% in the BETA measurements. In fractional terms these approximately correspond to the sources above our third flux density bin in Figure \ref{fig:ratios}, so the discrepancy is consistent with our full bandwidth continuum measurements. Serra et al.~(2015) remark that inaccuracies in models of the primary beams may be the cause, and such effects would certainly also be present in the data presented here. Calibration and deconvolution biases may also be present in both data sets, despite efforts to minimise them.
	
\begin{figure}
\centering
\includegraphics[width=\columnwidth]{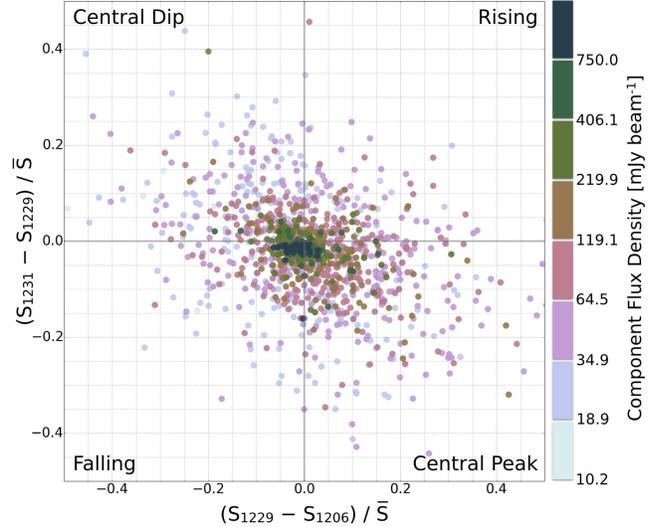}
\caption{The differences in the peak component flux densities for a pair of epochs expressed as a fraction of the mean value for that pair. The central epoch (SB 1229) is chosen as a reference epoch and the plot shows the SB 1231 to SB 1229 values against the SB 1229 to SB 1206--1207 values, i.e.~the first and second epoch are compared on the x-axis, and the second and third epoch are compared on the y-axis. The matched components are placed into logarithmically-spaced bins depending on their peak flux density value in the reference epoch, SB 1229, as indicated by the colour scale. The increased scatter due to the decreasing signal-to-noise ratio is clear.}
\label{fig:sdiffs}
\end{figure}

Epoch to epoch stability is demonstrated in Figure \ref{fig:sdiffs}. This shows the differences in the peak component flux density values for pairs of epochs, expressed as a fraction of the mean value for the pair. The x-axis compares the first and the second epoch and the y-axis compares the second and the third epoch. Again, the points are colour coded by their peak flux density, as measured in the second epoch (SB 1229). The increased spread due to decreasing signal-to-noise ratio is evident. Note that the catalogues exclude sources that have a peak flux density of less than 5$\sigma$, so a two epoch comparison of a source at the detection threshold for two observations of equal depth will by definition exhibit apparent variability at the 20\% level. The diagonal bias in Figure \ref{fig:sdiffs} results naturally from making this type of plot from noisy data. Making three measurements of a source of fixed brightness in the presence of independent Gaussian noise will result in a light curve that is either rising, falling, or peaks or dips in the centre. These four scenarios place the object in one of the marked quadrants on Figure \ref{fig:sdiffs}. For a large number of measurements the upper left and lower right (peaking or dipping) quadrants will contain twice as many points as the rising or falling quadrants. It is trivial to prove this with a simple Monte Carlo simulation.

The broadening of the distribution in the x-direction is likely due to the decreased effective depth of the first epoch. There is also an offset from the zero line in the y-direction, suggesting that a $\sim$2--3\% offset in the measured values is present for the final epoch. Examination of the calibrated spectra of PKS B1934$-$638 for these two epochs shows no such offset, suggesting that this is either due to a system-wide gain drift associated with SB 1231 that the phase-only self-calibration does not correct, or an image-plane effect. Element gain drifts will cause derived beamformer weights to become increasingly inaccurate over time, and this slight degradation in the flux accuracy may be a manifestation of such `ageing' of the weights. 

How does this $\sim$2--3\% flux scale offset compare to measurements made with other synthesis arrays? This offset is not anomalously damaging. Perley \& Butler (2013) claim that at L-band the VLA will exhibit a scatter in measured flux densities of better than 1\% solely due to the accuracy of the flux-scale transfer from the primary calibrator, rising to 3\% at higher frequencies, although they note that non-optimal calibration methods and observing conditions can degrade this accuracy significantly. The ATCA exhibits similar levels of accuracy from the flux-scale transfer, better than 3\% (J. Stevens, private communication). Croft et al. (2010; 2011) present a multi-epoch transient survey using the 42-element Allen Telescope Array, and note flux scale offsets of up to a factor of 2 between their epochs. They ascribe this to the significantly differing ($u$,$v$) plane coverages between some of their 60 second snapshot observations introducing variable levels of clean bias. The PAF beams appear to be stable enough to conduct astronomical measurements over timescales of order one week, and in any case it is likely that the beam weights will be refreshed or adjusted on shorter timescales for the full ASKAP system, and increasing the calibration of the beam gain amplitudes would likely remove the offset we identify here.

\subsection{Astrometry}
\label{sec:astrometry}

The uncertainty in the position of a radio source as measured from an interferometric image generally has two components. These are a statistical component that is related to the signal to noise ratio of the detection and the angular resolution of the instrument, and a systematic component that is associated with errors in the astrometric reference frame that are generally calibration related. The former can be understood analytically (Condon, 1997) whereas the magnitude of the latter component is typically determined by comparing the measured positions of significant numbers of sources with external measurements of high accuracy, for example a set of calibrator sources with accurate VLBI positions, at flux densities that are high enough to render the statistical component insignificant (e.g.~Condon et al., 1998; Prandoni et al., 2000; Bondi et al. 2003). Indeed one of the purposes of regular phase referencing observations of a strong calibrator that is close to the target field is to tie the astrometric reference frame of the target field to the position of the calibrator source, which is typically known to high accuracy.

As mentioned in Section \ref{sec:selfcal}, only a single calibration scan of PKS B1934$-$638 was performed for each of the twelve hour target observations. While an initial set of phase corrections will be made as part of the bandpass and flux scale calibration, this may be inadequate for two reasons: (i) PKS B1934$-$638 is approximately 25 degrees away from the centre of the mosaic, and (ii) complex gain variations over the course of the observation may result in astrometric errors that will then be frozen in by the self-calibration procedure.

We investigate the magnitude of both the statistical and systematic calibration uncertainties in our observations by matching source positions to those of the SUMSS catalogue, and correcting for the latter effect by shifting the coordinate frame of the BETA observations by the mean offset in RA and Dec on a per-epoch basis. The combined three-epoch images are formed from the re-aligned data. The mean offsets in (RA, Dec) in arcseconds for the three epochs in chronological order are ($-$2.3,$-$3.5), ($-$12.3,$-$3.1) and ($-$4.5,$-$3.9), where north and west are positive, all a small fraction of the (70$''$~$\times$~60$''$) synthesised beam width.

\begin{figure}
\centering
\includegraphics[width=\columnwidth]{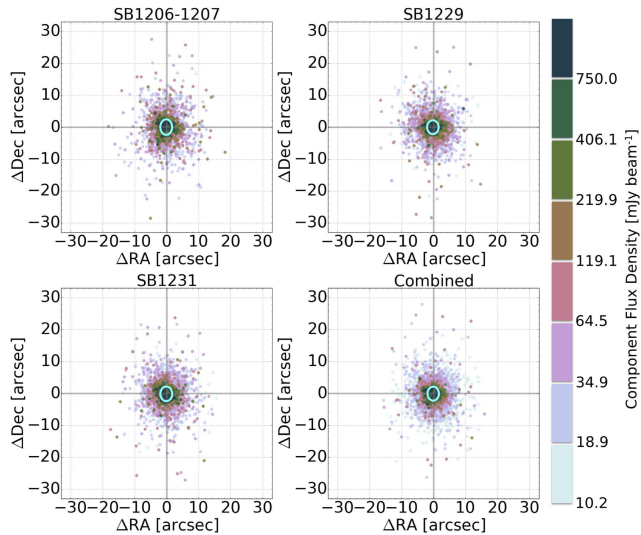}
\caption{Residual offsets in RA and Dec for BETA sources matched with counterparts from SUMSS, following the application of the bulk astrometric corrections as described in the text. Left to right, top to bottom, the panels show the three epochs and the final combined data, the number of matches sources being 1841, 2282, 2241 and 2927 respectively. The cyan ellipses show the 1$\sigma$ scatter in the measured offsets, representing the statistical uncertainties. The extent of these ellipses in (RA, Dec) in arcseconds are (3.7, 4.9), (3.6, 4.3), (3.6, 4.6) and (3.5, 4.4). Matched sources are colour coded by where their flux densities are placed in a set of logarithmically-spaced bins as per the colour scale. Note that the extent of the synthesised beam is 70$''$~$\times$~60$''$, i.e.~the approximate extent of the panels in the figure above. Elongation of the distribution in the north-south direction is likely to be related to the synthesised beam of SUMSS having an axial ratio of between 1.15 and 1.35 (longest in the north-south direction) over the declination range of our survey.}
\label{fig:astrometry}
\end{figure}

Figure \ref{fig:astrometry} shows the residual positional offsets in right ascension and declination between sources detected in the BETA data and their SUMSS counterparts, following the application of the corrections. The extent of the panels in these plots are equal to the minor axis of the BETA synthesised beam. As with Figure \ref{fig:sdiffs}, matched components are coloured by where their flux densities lie in a set of logarithmically-spaced bins as per the colour scale. The stronger sources have much tighter constraints on their positions as expected, an effect that also manifests itself via the shrinking of the cyan ellipse. The extent of this ellipse in RA and Dec is the 1$\sigma$ scatter in the residual offsets. It is largest for the noisiest first epoch (upper left panel) and smallest for the final combined mosaic (lower right panel). The extents of these ellipses are provided in the figure caption. Note also that there is only a single cluster of points around the (0,0) position for all mosaics, indicating that there are no significant beam-dependent astrometric errors. The extension of the distribution in the north-south direction is likely due to the elongation of the restoring beam in that direction, as well as that of the SUMSS synthesised beam, also elongated in the north-south direction and exhibiting axial ratios of between 1.15 and 1.35 over the declination range covered by our imaging. 

Following the positional corrections, the coordinate frame of the BETA observations is therefore tied to that of the SUMSS survey, which was in turn verified against that of NVSS. The errors in (RA, Dec) for the SUMSS survey are (1.5, 1.7) arcseconds (Mauch et al., 2003), and we can therefore expect a similar uncertainty in the BETA positions due to astrometric frame errors, i.e.~$<$3\% of the extent of a synthesised beam at worst. The 1$\sigma$ statistical uncertainty in the one-dimensional position of a radio component is approximately $\theta$/2$S$ where $\theta$ is the size of the restoring beam along that direction and $S$ is the signal to noise ratio of the detection. The positional accuracies of radio surveys are generally best checked by cross matching them with an external set of higher resolution or higher signal to noise observations such that the errors are dominated by those in the survey being verified. There are insufficient ATCA calibrators with high positional accuracy within our survey area to perform such a test. The fact that we are comparing the positions to an external data set with similar angular resolution and depth means that the statistical uncertainties will not obey the reciprocal relationship with $S$, hence the full band uncertainties do not improve on the sub-band uncertainties by a factor of two.

The calibration approach for the full ASKAP array will involve the use of a Global Sky Model (Cornwell et al., 2011). Ideally the minimum form that such a model would take would be akin to an existing calibrator list with accurate position measurements, but with the density of viable calibrator sources increased such that there was at least one in every 30 square degree ASKAP field. However the calibration model also assumes that the most recently determined set of calibration solutions is used to correct the instrumental state at the start of an observation. The stability exhibited over the course of this pilot continuum survey bodes well for this being a viable approach. 

\subsection{In-band spectral indices}
\label{sec:spectra}

\begin{figure*}
\begin{center}
\setlength{\unitlength}{1cm}
\begin{picture}(16,13)
\put(-1.8,-0.5){\includegraphics{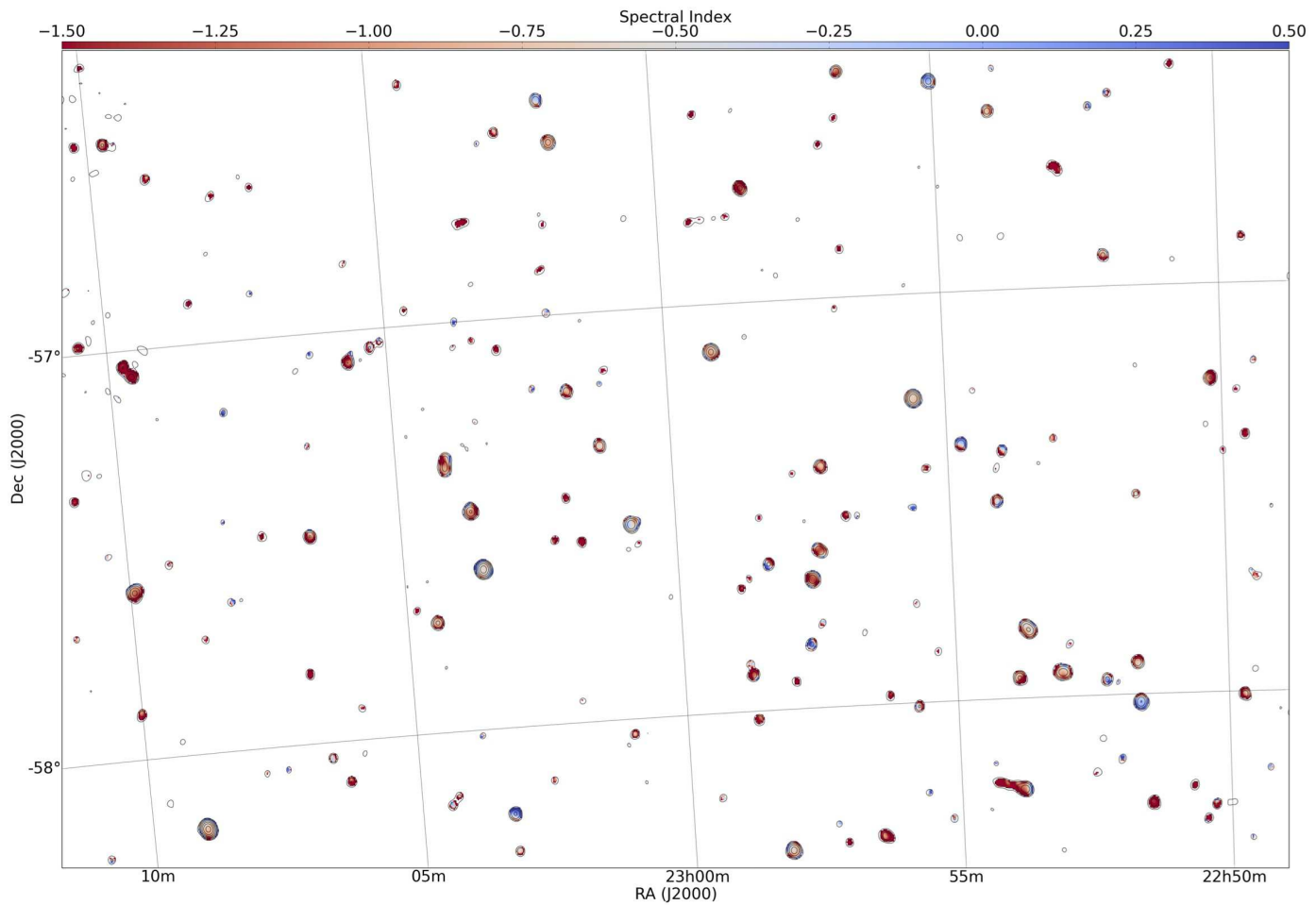}}
\end{picture}
\caption{A 3$^{\circ}$~$\times$~2$^{\circ}$ region of the combined ASKAP-BETA radio continuum mosaic, as indicated on Figure \ref{fig:mosaic}, with total intensity contours overlaid on the spectral index image. The base contour is 5~mJy beam$^{-1}$ and the levels increase in multiples of this according to the sequence (5$^{0}$, 5$^{0.5}$, 5$^{1.0}$, 5$^{1.5}$, $\ldots$).}
\label{fig:zoom}  
\end{center}
\end{figure*}

A four channel frequency cube from the sub-band mosaics was produced for the combined epoch data. Pixels with values less than 5 mJy beam$^{-1}$ were masked, corresponding to a cut at the 2--3$\sigma$ level depending on the band (see Figure \ref{fig:rmsplots}). A pixel-wise estimation of the spectral index ($\alpha$) was made by performing a linear fit to the log of the pixel values in log frequency space, for all directions where all four spectral points were unmasked, assuming $S \propto \nu^{\alpha}$ where $S$ is the flux density and $\nu$ is the frequency. The value of $\alpha$ that was determined for the non-masked pixels was then written to a FITS file to provide a spectral index map of the survey area. A 3~$\times$~2 degree sub image of the survey is shown in Figure \ref{fig:zoom} (corresponding to the boxed area highlighted on Figure \ref{fig:mosaic}) with the total intensity contours overlaid on the spectral index. The base contour is 5~mJy beam$^{-1}$ and the levels increase in multiples of this according to the sequence (5$^{0}$, 5$^{0.5}$, 5$^{1.0}$, 5$^{1.5}$, $\ldots$). A demonstration that significant astrometric errors across the band are not affecting the construction of the spectral index map is provided in Appendix B.

\begin{figure}
\centering
\includegraphics[width=\columnwidth]{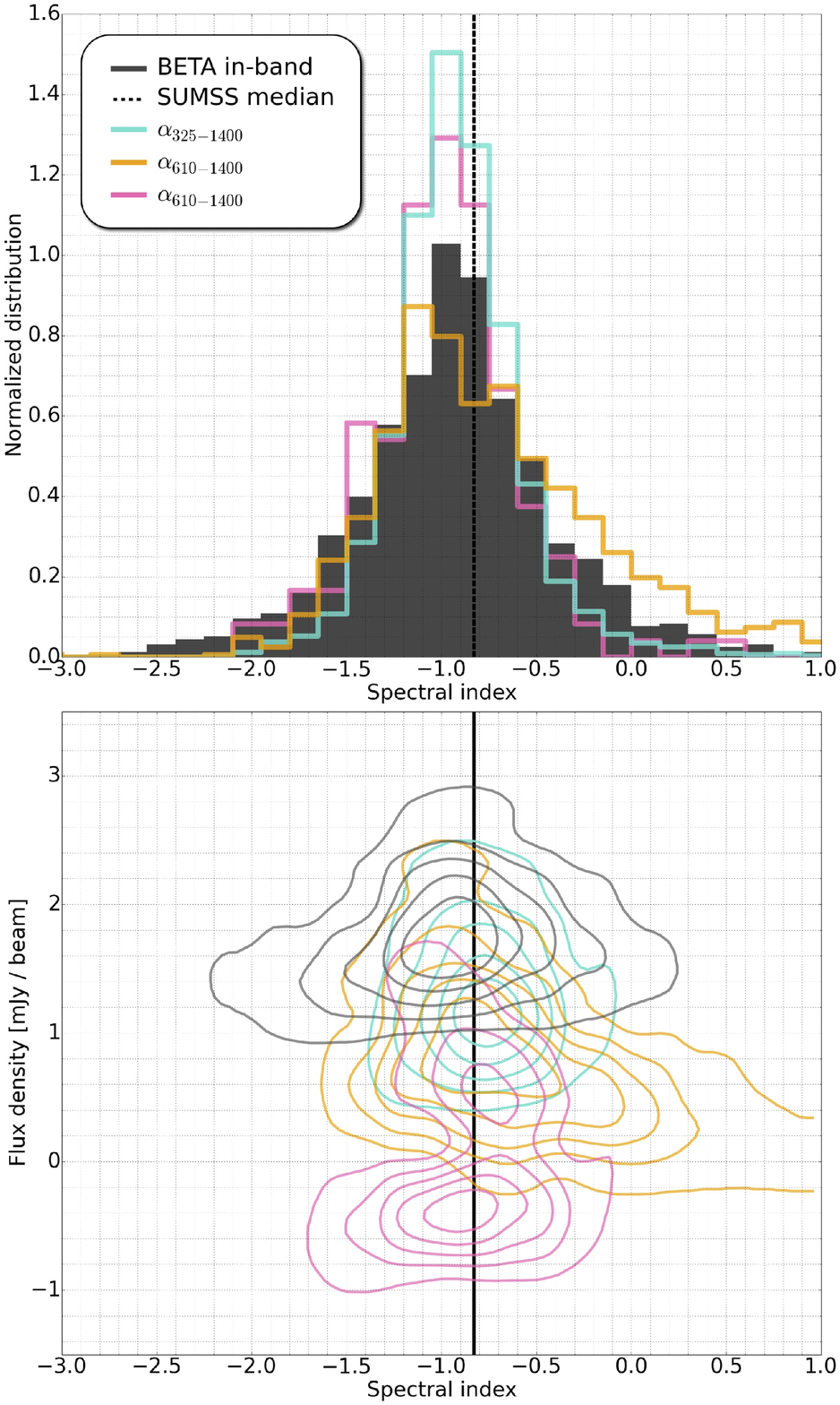}
\caption{Histograms of the distribution of spectral indices (defined as $S$~$\propto$~$\nu^{\alpha}$) from the BETA in-band measurements and the two-frequency external survey measurements are presented in the upper panel. The lower panel shows component flux density (interpolated to 863 MHz for the external measurements) as function of component spectral index. For clarity, contours are used to trace the normalized density of points in this plot, having values of (0.1, 0.3, 0.5, 0.7, 0.9). The vertical lines on both plots shows the median spectral index value of the SUMSS survey.}
\label{fig:alphahist}
\end{figure}

To verify the BETA in-band spectral index measurements we return to the component catalogues derived from the combined epoch data, in which the source finder will provide reliable single estimates of the component flux densities. The procedure for determining $\alpha$ is the same as the pixel-wise method, i.e.~a linear fit to the component flux density measurements against frequency in log-log space. The matching criteria were the same as those described in Section \ref{sec:astrometry} with the additional constraint that a component must have a matched detection in all four sub-bands. A normalized histogram of the BETA spectral index measurements is shown in grey in the upper panel of Figure \ref{fig:alphahist}. The median (with one standard deviation error) of the spectral index distribution is $-$0.92~$\pm$~0.57. Also shown on this figure is the overall median spectral index of $-$0.83 measured from the overlap region between SUMSS and NVSS (vertical line; Mauch et al., 2003). Additional normalized histograms show spectral index distributions as determined by cross-matching Giant Metrewave Radio Telescope (GMRT) observations at 325 and 610~MHz with VLA observations at 1400~MHz. Specifically, and in decreasing order of area and increasing order of depth, these are 325~MHz GMRT observations of the equatorial Galaxy And Mass Assembly fields (Mauch et al., 2013) cross matched with the FIRST survey (turquoise histogram), 610~MHz GMRT observations of the ELAIS-N1, ELAIS-N2, \emph{Spitzer} First Look Survey and Lockman Hole fields (Garn et al.,~2007; Garn et al.,~2008a; Garn et al.~2008b) cross-matched with the FIRST survey (orange histogram) and GMRT 610~MHz and VLA 1400~MHz observations of the VIMOS VLT Deep Survey field (Bondi et al.,~2003; Bondi et al.~2007; pink histogram). Summaries of these distributions can be found in Table \ref{tab:alphas}. The positive spectral index tail in the Garn et al. (2008a, 2008b) data is likely due to artifactual faint sources in their catalogue, the significant number of which results in many spurious associations with the higher frequency data (Ibar et al., 2009). The BETA in-band measurements agree with both the MOST-VLA and GMRT-VLA distributions within the scatter.

\begin{table}
\centering
\caption{Source spectral index ($\alpha$, where flux density $S \propto \nu^{\alpha}$ for frequency $\nu$) properties as derived from in-band BETA measurements and two-band measurements from existing radio surveys. The median and standard deviation values of the spectral index distributions are presented, corresponding to the measurements plotted in Figure \ref{fig:alphahist}. References for the external surveys used are provided in the text. This table lists the surveys in decreasing order of depth.}
\begin{tabular}{cccccc} \hline
$\nu_{1}$ & $\nu_{2}$ & $\alpha_{med}$ & $\sigma_{\alpha}$ & Array$_{1}$ & Array$_{2}$ \\
(MHz)     & (MHz)     &                &          &             &             \\ \hline
843       & 1400      & $-$0.83          & --       & MOST        & VLA (NVSS)  \\
711       & 1015      & $-$0.92          & 0.57     & BETA  & --          \\
325       & 1400      & $-$0.78          & 0.34     & GMRT        & VLA (FIRST) \\
610       & 1400      & $-$0.68          & 0.63     & GMRT        & VLA (FIRST) \\ 	
610       & 1400      & $-$0.84          & 0.42     & GMRT        & VLA (VIMOS)        \\ \hline
\end{tabular}
\label{tab:alphas}
\end{table}

A closer angular resolution match was the reason that the FIRST survey (5$''$) was chosen over NVSS (45$''$) for the GMRT (14--23.5$''$ at 325~MHz, 5$''$ at 610~MHz) cross matching in an effort to minimise resolution biases from contaminating the spectral index measurements, and integrated flux as opposed to peak flux density measurements were used for the same reason. We note that the median spectral indices we determine are in good agreement with the GMRT-NVSS measurements ($-$0.71~$\pm$~0.38) presented by Mauch et al.~(2013), and that approximately three quarters of the 325~MHz GMRT sources are marginally resolved or unresolved ($S_{peak}$/$S_{int}$~$>$~0.95).

The lower panel on Figure \ref{fig:alphahist} shows the spectral index distribution as a function of 863~MHz flux density. This is expressed as a contour plot of the normalized 2D histogram of the data points in the flux density against spectral index plane with contour values of (0.1, 0.3, 0.5, 0.7 and 0.9). Colour coding of the various surveys shown are consistent with those of the upper panel. Natural broadening of the distribution with decreasing flux density occurs due to the same signal to noise considerations that have been discussed previously. GMRT-VLA measurements are interpolated to 863~MHz. SUMSS values are not adjusted from their 843 MHz measurements.

Transitions in the typical spectral index at high flux densities occur due to the shift between flat spectrum quasars and steeper spectrum radio galaxies (at $>$100~mJy), more readily seen in higher frequency observations (e.g.~Massardi et al., 2011). At fainter levels Mauch et al.~(2003) report a shift towards flatter spectrum sources with decreasing flux density, likely due to the transition from steep spectrum Fanaroff-Riley Type-I radio galaxies (Fanaroff \& Riley, 1974) to lower luminosity flat spectrum AGN. Some studies (e.g.~Randall et al., 2012) report no significant shift in the typical source spectral index at $\sim$mJy levels.
  
The GMRT and VLA observations we have selected for comparison to the BETA data form a suitable tiered arrangement in terms of their depth and area, and the shallowest and deepest observations report steeper median spectral indices than the central observations. Insight into a possible shift in the typical radio source spectral index as a function of flux density is beyond the scope of this paper, and indeed is not feasible with the BETA data presented. 

A possibly significant consideration for any systematic errors in the BETA measurements is that the simple assumptions made about the primary beam (Section \ref{sec:imaging}) are inadequate. A model that assumed a primary beam that was narrower than the actual beam at the top end of the band (or did not faithfully capture a frequency scaling that differed significantly from the 
usual linear behaviour) would result in a systematic steepening of the spectral index of a source when measured using the sub-band mosaicking technique described above. We touch on this issue again in Appendix \ref{sec:singlebeams}. Beam formation and shape determination is a primary activity in commissioning of ASKAP, and it will be informative when this issue is revisited as part of future ASKAP observations employing shape-constrained primary beams.

\subsection{Transient and variability search}
\label{sec:variability}

The Variables and Slow Transients (VAST; Murphy et al., 2013) pipeline ingests a time series of images and performs a comparative analysis in order to detect variability via both image to image comparison and with reference to an external catalogue, in this case SUMSS. A full description of the pipeline is provided by Bell et al.~(2014). No significant variability was found across the week-long time baseline of the observations we present here; however when comparing the images to the SUMSS data, the pipeline reported a single detection. The identified source was J223825$-$511419 with a flux density of 24.1 mJy in the SUMSS catalogue and a flux density of 52.8 mJy in our combined image. The source is a candidate quasar, detected at 5, 8 and 20 GHz as part of the AT20G survey (Murphy et al., 2010), and is also a ROSAT X-ray source. Its position is marked on Figure \ref{fig:mosaic}.

In comparable work, Bannister et al. (2011) search for variable radio sources within archival Molonglo Observatory Synthesis Telescope images. The survey covered 2,776 deg$^{2}$ at an observing frequency of 843~MHz with 5$\sigma$ sensitivity $>$14~mJy~beam$^{-1}$.
Over their survey area they report the detection of 53 variable sources, which implies a surface density of variables on the sky of $0.02$~deg$^{-2}$. The work by Bannister et al. (2011) focused on a variety of variability timescales from 24 hours to 20 years.
By comparing these BETA observations with SUMSS we probe a timescale of $\sim$ 17.5 years only.
Scaling the surface density reported by Bannister et al. (2011) we expect $\sim$3 variable sources in this study.
Accounting for differences in timescales probed we conclude that these numbers are in good agreement.
A cross-check of the variable source found in this survey with the list of sources reported by Bannister et al. (2011) revealed that they did not identify it as being significantly variable.

\subsection{Source catalogue}
\label{sec:catalogue}

The {\tt PyBDSM} source finder was used to extract a component catalogue from the deep mosaic image formed from a combination of all epochs and sub-bands. Components were fit to islands of emission that had a peak brightness of $>$5$\sigma$ and an island boundary threshold of $>$3$\sigma$, where $\sigma$ is the local estimate of the background noise level. Component spectral indices are assigned by matching positions at which spectral indices were successfully fit (Section \ref{sec:spectra}). Following the excision of some spurious detections at the noisy edge of the mosaic the final catalogue contains 3,722 components, 1,037 of which have in-band spectral index measurements. Ten random entries are extracted from the catalogue and presented in Table \ref{tab:catalogue} to demonstrate the structure. The columns are defined as follows:\\
\noindent
(1) Identifier for the component formed from its HHMMSS.SS+/-DDMMSS.SS right ascension and declination position in J2000 coordinates.\\
\noindent
(2) Right ascension of the component in degrees.\\
\noindent
(3) Declination of the component in degrees.\\
\noindent
(4) The 1$\sigma$ uncertainty in right ascension in arcseconds.\\
\noindent
(5) The 1$\sigma$ uncertainty in declination in arcseconds.\\
\noindent
(6) Integrated flux density of the component in mJy.\\
\noindent
(7) The 1$\sigma$ uncertainty in the integrated flux density of the component in mJy.\\
\noindent
(8) Peak flux density of the component in mJy beam$^{-1}$.\\
\noindent
(9) The 1$\sigma$ uncertainty in the peak flux density of the component in mJy beam$^{-1}$.\\
\noindent
(10) Estimate of the local RMS noise at the position of the component in mJy beam$^{-1}$.\\
\noindent  
(11) Spectral index ($\alpha$) of the component formed via a fit to the component flux densities across the four sub-bands. See Section \ref{sec:spectra} for further details.\\
\noindent  
(12) Deconvolved major axis size of the Gaussian fitted to the component in arcseconds. A value of zero indicates that the source is unresolved.\\
\noindent
(13) The 1$\sigma$ uncertainty in the major axis of the component in arcseconds.\\
\noindent
(14) Deconvolved minor axis size of the Gaussian fitted to the component in arcseconds. A value of zero indicates that the source is unresolved.\\
\noindent
(15) The 1$\sigma$ uncertainty in the minor axis of the component in arcseconds.\\
\noindent
(16) Position angle measured east of north of the Gaussian fitted to the component in degrees.\\
\noindent
(17) The 1$\sigma$ uncertainty in the position angle of the component in degrees.\\
\noindent
(18) Zero-indexed unique identifier for the component.\\
\noindent
(19) Zero-indexed unique identifier for the source.\\
\noindent
(20) Zero-indexed unique identifier for the island.\\

Note that the values quoted in columns (4) and (5) are derived from the uncertainty in the component fitting and are thus sensitive only to the statistical component discussed in Section \ref{sec:astrometry}. A more rigorous estimation of the positional errors can be achieved by the quadrature sum of the listed values and the astrometric uncertainties that are tied to those of SUMSS (see Section \ref{sec:astrometry}). The complete catalogue is available online as supplementary material.

\begin{table*}
\begin{minipage}{170mm}
\centering
\begin{tabular}{ccccccccccc} \hline
	\multicolumn{1}{c}{ID} &
	\multicolumn{1}{c}{R.A.} &
	\multicolumn{1}{c}{Decl.} &
	\multicolumn{1}{c}{$\sigma_{\trm{\tiny R.A.}}$} &
	\multicolumn{1}{c}{$\sigma_{\trm{\tiny Decl.}}$} &
	\multicolumn{1}{c}{$S_{\trm{\tiny int}}$} &
	\multicolumn{1}{c}{$\sigma_{S_\trm{\tiny int}}$} &
	\multicolumn{1}{c}{$S_{\trm{\tiny peak}}$} &
	\multicolumn{1}{c}{$\sigma_{S_\trm{\tiny peak}}$} &
	\multicolumn{1}{c}{Local RMS} \\
	&
	\multicolumn{1}{c}{[deg]} &
	\multicolumn{1}{c}{[deg]} &
	\multicolumn{1}{c}{[arcsec]} &
	\multicolumn{1}{c}{[arcsec]} &
	\multicolumn{1}{c}{(mJy)} &
	\multicolumn{1}{c}{(mJy)} &
	\multicolumn{1}{c}{(mJy~b$^{-1}$)} &
	\multicolumn{1}{c}{(mJy~b$^{-1}$)} &
	\multicolumn{1}{c}{(mJy~b$^{-1}$)} \\
	\multicolumn{1}{c}{(1)} &
	\multicolumn{1}{c}{(2)} &
	\multicolumn{1}{c}{(3)} &
	\multicolumn{1}{c}{(4)} &
	\multicolumn{1}{c}{(5)} &
	\multicolumn{1}{c}{(6)} &
	\multicolumn{1}{c}{(7)} &
	\multicolumn{1}{c}{(8)} &
	\multicolumn{1}{c}{(9)} &
	\multicolumn{1}{c}{(10)} \\ \hline
J225457.52$-$500138.10&  343.7397&  $-$50.0273&    0.11&    0.16&   283.57&     2.01&   254.72&     1.20&     1.17& \\
J225303.43$-$575625.37&  343.2643&  $-$57.9404&    0.55&    0.41&   149.48&     1.99&    94.69&     1.32&     1.23& \\
J222640.45$-$580424.37&  336.6686&  $-$58.0734&    8.60&    5.05&    30.84&     2.48&    12.43&     1.83&     1.75& \\
J231720.60$-$593405.91&  349.3359&  $-$59.5683&    1.08&    1.23&    21.72&     1.85&    23.62&     1.04&     1.06& \\
J225656.90$-$502229.66&  344.2371&  $-$50.3749&    2.14&    4.31&    12.01&     2.23&    11.66&     1.27&     1.29& \\
J224937.76$-$582124.10&  342.4074&  $-$58.3567&    5.72&    6.21&     8.32&     2.51&     7.45&     1.49&     1.46& \\
J230310.41$-$615333.05&  345.7934&  $-$61.8925&    3.51&    3.29&     7.72&     2.04&     8.63&     1.14&     1.17& \\
J230627.57$-$531436.35&  346.6149&  $-$53.2434&    4.72&    7.19&     8.63&     2.24&     7.25&     1.35&     1.32 \\
J223153.75$-$543914.32&  337.9740&  $-$54.6540&    4.84&    5.78&     7.12&     2.33&     7.48&     1.29&     1.34 \\
J224253.27$-$584415.57&  340.7220&  $-$58.7377&    3.37&    7.26&     6.45&     2.39&     7.21&     1.27&     1.36 \\ \hline
	\multicolumn{10}{c}{} \\ \hline
	\multicolumn{1}{c}{$\alpha$} &
	\multicolumn{1}{c}{$\Theta_{maj}$} &
	\multicolumn{1}{c}{$\sigma_{\Theta_{maj}}$} &
	\multicolumn{1}{c}{$\Theta_{min}$} &
	\multicolumn{1}{c}{$\sigma_{\Theta_{min}}$} &
	\multicolumn{1}{c}{PA} &
	\multicolumn{1}{c}{$\sigma_{PA}$} &
	\multicolumn{1}{c}{ID$_{g}$} &
	\multicolumn{1}{c}{ID$_{s}$} &
	\multicolumn{1}{c}{ID$_{i}$} \\
	\multicolumn{1}{c}{} &
	\multicolumn{1}{c}{[arcsec]} & 
	\multicolumn{1}{c}{[arcsec]} &
	\multicolumn{1}{c}{[arcsec]} &
	\multicolumn{1}{c}{[arcsec]} &
	\multicolumn{1}{c}{[deg]} &
	\multicolumn{1}{c}{[deg]} &
	&
	&\\
	\multicolumn{1}{c}{(11)} &
	\multicolumn{1}{c}{(12)} &
	\multicolumn{1}{c}{(13)} &
	\multicolumn{1}{c}{(14)} &
	\multicolumn{1}{c}{(15)} &
	\multicolumn{1}{c}{(16)} &
	\multicolumn{1}{c}{(17)} &
	\multicolumn{1}{c}{(18)} &
	\multicolumn{1}{c}{(19)} &
	\multicolumn{1}{c}{(20)}\\ \hline
   $-$0.41&    25.40&     0.37&    17.56&     0.27&   28.48&  178.70&   2359&   2205&   2249\\
   $-$0.77&    64.90&     1.31&    26.85&     0.98&   85.03&    1.54&   2615&   2445&   2498\\
       -&   109.86&    20.95&    44.74&    10.72&   79.12&   21.84&   4079&   3776&   3903\\
   $-$1.07&     0.00&     2.91&     0.00&     2.53&    0.00&   19.11&   1347&   1292&   1292\\
       -&     0.00&    10.16&     0.00&     5.01&    0.00&   11.50&   2228&   2089&   2128\\
       -&    37.05&    16.11&     0.00&    11.67&   58.80&   51.97&   2829&   2639&   2703\\
       -&     0.00&     8.33&     0.00&     7.69&    0.00&  134.18&   2172&   2035&   2075\\
       -&    47.41&    17.85&     0.00&     9.60&  143.22&   22.22&   1695&   1601&   1621\\
       -&     0.00&    16.19&     0.00&     7.29&    0.00&   23.13&   3884&   3600&   3714\\
       -&     0.00&    17.68&     0.00&     6.55&    0.00&   17.77&   3226&   2998&   3081\\
\hline
\end{tabular}
\caption{A subset of ten random components from the final catalogue, presented here in order to demonstrate the table structure. Columns (6), (7), (8) and (9) are computed at the band centre reference frequency, 863 MHz. Please refer to the text for a detailed description of each column. The full catalogue is available online as supplementary material.}\label{tab:catalogue}
\end{minipage}
\end{table*}

\subsection{Differential source counts at 863~MHz}
\label{sec:sourcecounts}

\begin{figure}
\centering
\includegraphics[width=\columnwidth]{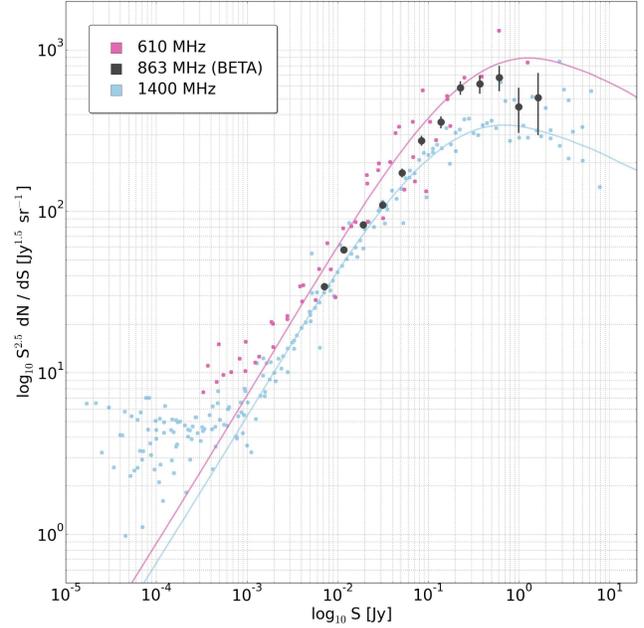}
\caption{Differential source counts normalized to a Euclidean universe with observationally derived counts at 610 MHz (pink points) and 1400 MHz (blue points), with the 863 MHz counts derived from the BETA data shown in black (see also Table \ref{tab:sourcecounts}). The observed counts shown are from various authors, as tabulated by de Zotti et al.~(2010). Error bars have been omitted from the 610 and 1400~MHz data for clarity. The solid lines show the corresponding AGN count models by Massardi et al.~(2010).}
\label{fig:sourcecounts}
\end{figure}

Figure \ref{fig:sourcecounts} shows the Euclidean-normalized differential source counts at 863 MHz derived from the BETA data. Error bars are Poissonian. The source counts are determined from the final catalogue in logarithmically-spaced flux density bins. The visibility area, i.e. the area over which the survey is capable of detecting sources with brightnesses exceeding the bin centre, is calculated from the estimated background noise map produced by the source finder. The lowest flux bin has a visibility area of 64 square degrees, thus the contribution to the count uncertainties introduced by field to field variations are negligible (Heywood, Jarvis \& Condon, 2013). The source counts are presented in Table \ref{tab:sourcecounts}, and are in good agreement with the 843 MHz counts determined from SUMSS (Mauch et al., 2003). Also shown on the figure are observed differential source counts at 610~MHz (pink points) and 1400~MHz (blue points), taken from the review paper by de Zotti et al.~(2010). The solid lines on the plot show the AGN population models of Massardi et al.~(2010). The BETA measurements are clearly AGN-dominated, with the deviation of the measurements from the 1400~MHz model below $\sim$1~mJy being due to the aforementioned increasing contribution of star forming galaxies to the source counts.

\begin{table}
\centering
\caption{Raw and Euclidean-normalized differential source counts for the final catalogue.}
\begin{tabular}{ccccc} \hline
$S_{\mathrm{bin}}$ & $A$     & $N$ & $S^{2.5} dN/dS$ & Error \\
(mJy)     & (sq.deg.) &      & (Jy$^{1.5}$~sr$^{-1}$) &(Jy$^{1.5}$~sr$^{-1}$) \\ \hline
6.8 & 63.6 & 568 & 34.2 & 1.4 \\
11.2 & 118.5 & 854 & 57.9 & 2.0 \\
18.3 & 148.4 & 728 & 82.6 & 3.1 \\
30.0 & 154.8 & 481 & 109.7 & 5.0 \\
49.2 & 155.8 & 364 & 172.9 & 9.1 \\
80.5 & 155.8 & 275 & 273.8 & 16.5 \\
131.9 & 155.8 & 171 & 356.8 & 27.3 \\
216.0 & 155.8 & 133 & 581.8 & 50.4 \\
353.8 & 155.8 & 67 & 614.4 & 75.1 \\
579.5 & 155.8 & 35 & 672.7 & 113.7 \\
949.2 & 155.8 & 11 & 443.2 & 133.6 \\
1554.7 & 155.8 & 6 & 506.7 & 206.9 \\
 \hline
\end{tabular}
\label{tab:sourcecounts}
\end{table}

\section{Conclusion}
\label{sec:conclusion}

The results presented in this paper demonstrate the viability and clear potential for using PAF receivers to rapidly and accurately conduct broadband continuum imaging of the radio sky. BETA consists of 6 of the 36 ASKAP antennas, equipped with prototype Mark I PAF receivers, and we have presented the results of a pilot continuum imaging survey conducted with this instrument at 711--1015 MHz. The sky coverage of the survey is approximately 150 square degrees in the constellation of Tucana, achieved by forming nine simultaneous PAF beams on the sky and mechanically repointing the telescope on a five-minute duty cycle through an appropriately spaced grid of twelve sky positions. Approximately two-thirds of the \emph{Spitzer} South Pole Telescope Deep Field is covered. The survey area was observed in three 12-hour runs over the course of a week with the goal of verifying both the on-sky performance and the stability of the PAFs.

Bandpass and flux density corrections were derived from a per-beam scan of PKS B1934$-$638 associated with each of the three epochs, and (self-)calibration and imaging was achieved with a fully automated pipeline based on standard packages. The principal data products from the survey are wide field radio images of the survey area in 4~$\times$~76~MHz sub-bands for each of the three epochs, as well as a single image formed using the full bandwidth. Additionally the three epochs have been combined into a single deep data set, for which sub-band and full-band mosaics were also generated. Source component lists were derived from each of these 20 images using the {\tt PyBDSM} source finder.

The survey mode performance and stability of the BETA telescope and its compound beams were verified by comparing the three-epoch data with external measurements, principally from the SUMSS survey which is well matched to the BETA data in both angular resolution, depth and observing frequency. The positions and flux densities of almost three thousand radio sources were measured with BETA and verified to be as accurate as those from existing radio survey measurements.

The major feature of our results that supersedes those from SUMSS comes from the use of the 35\% fractional bandwidth of the observations to determine in-band spectral index measurements. A spectral index map of the survey area was produced for emission above 5~mJy beam$^{-1}$, and the signal to noise ratio of 1,037 of the 3,722 components in the final catalogue was sufficient for spectral index measurements to be assigned to them. The spectral index distribution was found to be in good agreement with those derived from dual-frequency measurements made by interpolating GMRT and MOST data at 325, 610 and 843 MHz with VLA data at 1400 MHz. We also used the final catalogue to measure the differential source counts at 863 MHz. A search for variability and transient events between the three epochs, as well as thorough comparison to the SUMSS survey, reveals one significantly variable source: a candidate quasar for which the $\sim$850~MHz flux density has approximately doubled on a 17.5 year time scale.

\section*{ACKNOWLEDGEMENTS} 

We thank the anonymous referee and the MNRAS editorial staff for their comments on this paper. The Australian SKA Pathfinder is part of the Australia Telescope National Facility which is managed by CSIRO. Operation of ASKAP is funded by the Australian Government with support from the National Collaborative Research Infrastructure Strategy. Establishment of the Murchison Radio-astronomy Observatory was funded by the Australian Government and the Government of Western Australia. ASKAP uses advanced supercomputing resources at the Pawsey Supercomputing Centre. We acknowledge the Wajarri Yamatji people as the traditional owners of the Observatory site. This work was supported by resources provided by the Pawsey Supercomputing Centre with funding from the Australian Government and the Government of Western Australia. Parts of this research were conducted by the Australian Research Council Centre of Excellence for All-sky Astrophysics (CAASTRO), through project number CE110001020. This research made use of Montage. It is funded by the National Science Foundation under Grant Number ACI-1440620, and was previously funded by the National Aeronautics and Space Administration's Earth Science Technology Office, Computation Technologies Project, under Cooperative Agreement Number NCC5-626 between NASA and the California Institute of Technology. This research has made use of NASA's Astrophysics Data System. This research has made use of the NASA/IPAC Extragalactic Database (NED) which is operated by the Jet Propulsion Laboratory, California Institute of Technology, under contract with the National Aeronautics and Space Administration. IH acknowledges many useful discussions with Jan Noordam, Rick Perley and Oleg Smirnov.

\appendix
\section{Examination of the individual beam patterns}
\label{sec:singlebeams}

\begin{figure*}
\begin{center}
\setlength{\unitlength}{1cm}
\begin{picture}(16,11.9)
\put(1.68,-0.2){\includegraphics{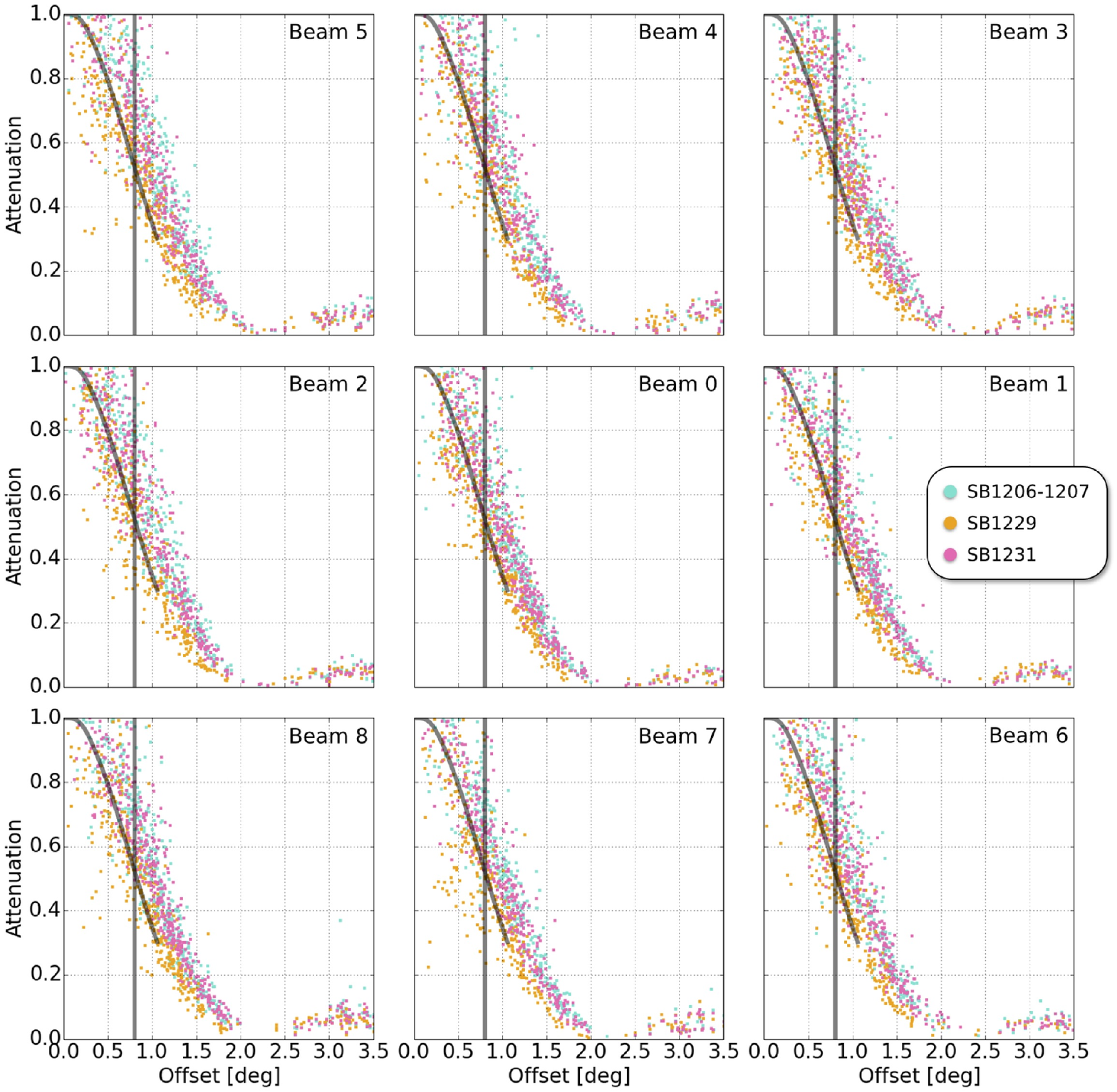}}
\end{picture}
\caption{Estimates of the radial profiles of the nine formed beams made by measuring the ratio of the uncorrected flux density measurement of the components in an 825.6~MHz BETA image to the intrinsic 843~MHz SUMSS measurement. The main lobe of each of the BETA primary beams is evident, as is the first sidelobe. The points are colour coded by epoch as illustrated on the key. The line following the points is the assumed primary beam shape used when corrected the images and forming the mosaic. The vertical line is the maximum separation of any pair of beams in the mosaic, crossing the model at the half power point.}
\label{fig:singlebeams}  
\end{center}
\end{figure*}

ASKAP is a survey telescope and its primary mode of operation will involve filling the $\sim$30 square degree field of view with an overlapping arrangement of beams. The base image product generated by the telescope will cover the full field of view, images created from individual beams are not envisaged to be a routinely produced data product. The accuracy of the multi-beam image is however critically dependent on the behaviour of the beams that are used to form it. Significant problems with individual beams result in obvious defects in the mosaic, and the accuracy and repeatability of the multi-beam images that we have demonstrated in this paper demonstrate consistency in the individual electronic beams that were used to produce them. However for completeness, in this section we present an examination of the properties of the nine simultaneous beams that were deployed on the BETA hardware.

A crude measurement of the total intensity (Stokes I) beam shape can be made by taking the images formed from each beam prior to the primary beam correction, and measuring the apparent to intrinsic brightness ratios of the sources as a function of distance from the nominal beam centre. The apparent flux density measurements were made in one of the four BETA sub-bands centred on 825.6~MHz. This was selected as it is closest to the SUMSS frequency of 843~MHz, which we use to provide the estimate of the intrinsic source flux density. No spectral correction is applied when measuring the ratio. Component matches are subject to all the criteria described Section \ref{sec:photometry}.

The apparent to intrinsic component flux density ratios as a function of distance from the nominal beam centre are plotted as the points on Figure \ref{fig:singlebeams}, for all nine beams in the 3~$\times$~3 beam footprint, laid out in their relative arrangement on the sky. The main lobe is evident, as is the first sidelobe. The points are colour coded according to their epoch, as indicated in the key. No flux density cut has been made in order to maximise the number of measurements on the plot. However note that a hypothetical source detected at 5$\sigma$ in both surveys could have an uncertainty in this ratio of up to $\sim$30\%, and such signal to noise considerations in both surveys result in the significant scatter in the plot. The line that traces the distribution of the points is the assumed primary beam model used to correct the images in this sub-band. It terminates at the 30\% level, which is where the images were cut when forming the mosaic (Section \ref{sec:imaging}). The vertical line shows the \emph{maximum} separation between a pair of beams in the full mosaic, crossing the beam model at the half power point.

No fit has been made between the assumed model and the points on this image, however the assumed beam model traces well the distribution of points, particularly in the region where errors would be most significant, i.e.~to the left of the vertical line. The model deviates from the points most as the beam gain lowers, however this is not uncommon, both in instruments with electronically formed beams (e.g.~Loi et al., 2015), and those with traditional single pixel feeds. Figure \ref{fig:singlebeams} demonstrates that there are no major beam to beam differences in the regions of interest, and the formation of mosaics with a single frequency-dependent Airy pattern model of the main lobe provides astronomical imaging performance that is as accurate as any existing comparable observations, at least in total intensity, as shown in Sections \ref{sec:photometry} and \ref{sec:astrometry}.

An examination of the frequency dependence of the beam patterns would be informative, particularly given that the median spectral index measured by the BETA observations appears to be slightly redder than those derived from other two-band, two-observation studies. As remarked in Section \ref{sec:spectra}, if the primary beam scales with frequency in a non-linear way then  this could bias the spectral measurements. The technique used to probe the beam shape presented here cannot be reliably extended to the frequency domain as one would have to assume a typical source spectral index in order to scale the SUMSS frequency, thus introducing a degeneracy with the very effect we would be attempting to constrain. The flux density of an $\alpha$~=~$-$0.8 source varies by approximately 30\% between 711 and 1015~MHz. We defer attempts to investigate this further to a forthcoming paper (McConnell et al., in prep.) which will include the results of an observing campaign to holographically measure BETA's formed beams.

Theoretical models of PAF beams suggest that the main lobes of the off-axis beams will exhibit small coma-like aberrations in their outer regions. The approach used to generate Figure \ref{fig:singlebeams} essentially azimuthally averages the beam profile. A two dimensional analogue of this plot does not have the necessary density of points to reliably reveal any azimuthal asymmetries, however there is apparent increased scatter in the measurements in the corner beams (e.g. beams 3 and 5) when compared to the central beam (beam 0) that could be a result of this effect.

Finally, although there is significant overlap between the points measured from the three epochs, it does appear that the central epoch (SB1229) exhibits a beam that is slightly narrower than the other epochs. As mentioned in Section \ref{sec:photometry}, the shape of an electronically formed beam can potentially be modulated by drifts in the electronic gains of the composite elements. Any distinct colour banding seen in a plot such as Figure \ref{fig:singlebeams} may be a manifestation of this.

\section{Astrometry across the band}

\begin{figure*}
\begin{center}
\setlength{\unitlength}{1cm}
\begin{picture}(16,10.7)	
\put(2.22,-0.2){\includegraphics{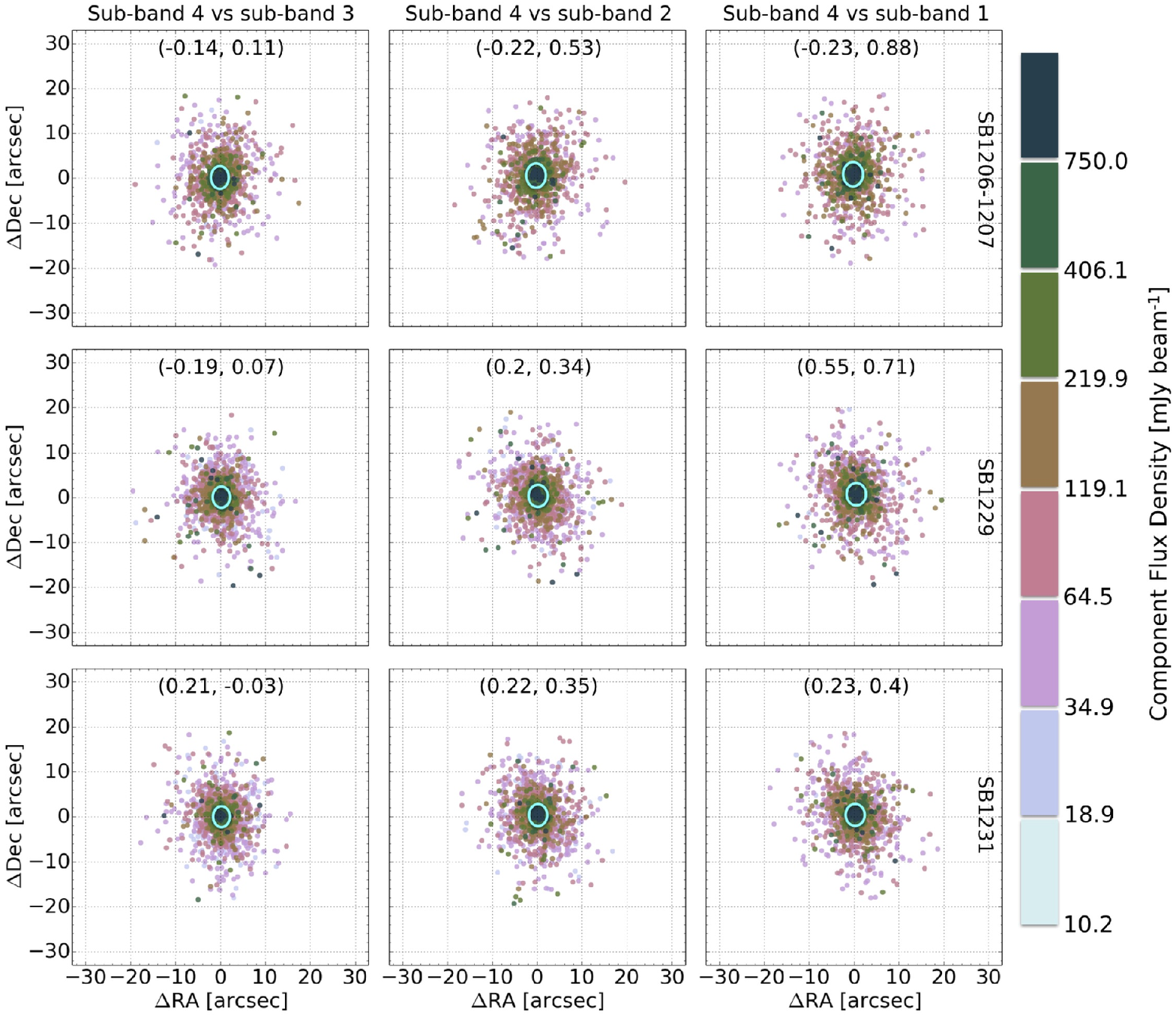}}
\end{picture}
\caption{Measured positional differences between components measured in sub-band 4 (the highest frequency sub-band) and matched components from the lower three sub-bands, as indicated at the top of each column. The rows show the distributions measured in each of the three epochs. As per Figure \ref{fig:astrometry} the points are colour coded by where the flux density of the sub-band 4 component lies in the set of logarithmically-spaced flux density bins as indicated by the legend. The cyan ellipses show the 1$\sigma$ scatter in the measured offsets and are centred on the mean offset positions, the exact values of which are provided on each panel.}
\label{fig:band_astrometry}  
\end{center}
\end{figure*}

A check of the positional accuracy of the sources in our images is made in Section \ref{sec:astrometry}; however, this was done in terms of the full bandwidth imaging. Here we present a measurement of the positional performance across the band in order to rule out the possibility of significant frequency dependent position shifts introducing gradients into the spectral index map (the spectral index values in the catalogue would not be affected). We measure this effect by taking the source catalogue associated with the highest frequency sub-band and cross matching the components with those found in the lower three sub-bands. This comparison is made for all three epochs, repeating the method used to make Figure \ref{fig:astrometry}, and the results are presented in Figure \ref{fig:band_astrometry}.

The points are once again colour coded by where the (highest frequency) flux density of the components lie in a set of logarithmically-spaced flux density bins as per the plot legend. No significant trend is seen as a function of frequency or epoch; the scatter introduced by signal to noise considerations is the dominant effect. Sub band images should have factor of two worse noise than full band images but scatter is comparable to the BETA vs SUMSS scatter, likely due to the contribution to the measured positional offsets that arises due to the SUMSS survey having comparable angular resolution and depth to our observations, as discussed in Section \ref{sec:astrometry}. The mean measured offsets (provided on the relevant panels of Figure \ref{fig:band_astrometry}) point to a self-consistent astrometric reference frame across the band. There is an increasing mean declination offset with decreasing frequency, e.g.~rising from 0.11$''$ to 0.88$''$ in SB1206-1207. Repeating the measurements using randomly drawn subsets consisting of 800 sources (roughly half the size of the per-sub-band catalogues) produces a distribution of mean offsets that is scattered around zero, however we cannot rule out some residual phase error across the band that remains due to calibration deficiencies. In any case, the magnitude of these offsets in comparison to the angular resolution of the survey is negligible.

\end{document}